\documentclass[times, twocolumn,trackchanges]{aastex701}
\usepackage{graphicx}
\usepackage{amssymb}
\usepackage{mathrsfs} 
\usepackage{amsmath}
\usepackage{natbib}
\usepackage[english]{babel}
\usepackage{url}
\usepackage{xcolor}
\usepackage{xspace}
\usepackage{subcaption}
\usepackage{CJK}

\newcommand{\redback}{\textsc{Redback}\xspace}

\newcommand{\scipy}{\textsc{SciPy}\xspace}
\newcommand{\mat}{\textsc{Matplotlib}\xspace}
\newcommand{\numpy}{\textsc{NumPy}\xspace}
\newcommand{\astropy}{\textsc{Astropy}\xspace}
\newcommand{\pyt}{\textsc{PyTorch}\xspace}
\newcommand{\sncosmo}{\textsc{SNCosmo}\xspace}
\newcommand{\gla}{\textsc{Glasflow}\xspace}
\newcommand{\nflows}{\textsc{nflows}\xspace}
\newcommand{\genova}{\textsc{Genova}\xspace}
\newcommand{\photutils}{\textsc{Photutils}\xspace}
\newcommand{\sfft}{\textsc{SFFT}\xspace}

\newcommand{\nichollbns}{N21\xspace}
\newcommand{\kasen}{K17\xspace}

\newcommand{\bns}{\texttt{nicholl\_bns}\xspace}

\newcommand{\ang}{\mbox{\AA}\xspace}

\newcommand{\flux}{\ensuremath{\mathrm{erg}\,\mathrm{s}^{-1}\,\mathrm{cm}^{-2}\,\mbox{\AA}^{-1}}\xspace}

\graphicspath{{./}{figures/}}

\begin{document}
\begin{CJK*}{UTF8}{gbsn}
\title{Probabilistic kilonova prediction from gravitational wave inferred binary neutron star parameters}

\author[0000-0001-7950-0531]{Xiao-Fei Dong}
\affiliation{SUPA, School of Physics and Astronomy, University of Glasgow, Glasgow G12 8QQ, UK}
\email{x.dong.2@research.gla.ac.uk}

\author[0000-0002-1977-0019]{Ik Siong Heng}
\affiliation{SUPA, School of Physics and Astronomy, University of Glasgow, Glasgow G12 8QQ, UK}
\email{Ik.Heng@glasgow.ac.uk}

\author[0000-0001-5169-4143]{Gavin P. Lamb}
\affiliation{Astrophysics Research Institute, Liverpool John Moores University, IC2 Liverpool Science Park, 146 Brownlow Hill, Liverpool L3 5RF, UK}
\email{G.P.Lamb@ljmu.ac.uk}

\author[0000-0001-7488-5022]{Chris Messenger}
\affiliation{SUPA, School of Physics and Astronomy, University of Glasgow, Glasgow G12 8QQ, UK}
\email{Christopher.Messenger@glasgow.ac.uk}

\author[0000-0003-2700-1030]{Nikhil Sarin}
\affiliation{Kavli Institute for Cosmology, University of Cambridge, Madingley Road, CB3 0HA, UK}
\affiliation{Institute of Astronomy, University of Cambridge, Madingley Road, CB3 0HA, UK}
\email{nsarin.astro@gmail.com}

\author[0009-0000-2285-8188]{Benjamin Rayson}
\affiliation{School of Physics and Astronomy, University of Leicester, University Road, Leicester LE1 7RH, UK}
\email{br155@leicester.ac.uk}

\author[0000-0003-3274-6336]{Nial R. Tanvir}
\affiliation{School of Physics and Astronomy, University of Leicester, University Road, Leicester LE1 7RH, UK}
\email{nrt3@le.ac.uk}

\author[0000-0002-2364-2191]{Jessica Irwin}
\affiliation{Institute for Gravitational and Subatomic Physics (GRASP), Utrecht University, Princetonplein 1, 3584 CC Utrecht, The Netherlands}
\email{j.irwin.1@research.gla.ac.uk}

\author[0009-0005-5527-7749]{Thomas Wallace}
\affiliation{SUPA, School of Physics and Astronomy, University of Glasgow, Glasgow G12 8QQ, UK}
\email{t.wallace.1@research.gla.ac.uk}

\author{Skye Rosetti}
\affiliation{Healey Close, Leicester, Leicestershire, LE4 2DH, UK}
\email{skye.rosetti@gmail.com}

\author[0000-0002-2281-2785]{Bo Milvang-Jensen}
\affiliation{Niels Bohr Institute, University of Copenhagen, Jagtvej 128, 2200
Copenhagen N, Denmark}
\affiliation{Cosmic Dawn Center (DAWN)}
\email{milvang@astro.ku.dk}

\author[0000-0001-7821-9369]{Andrew J. Levan}
\affiliation{Department of Astrophysics/IMAPP, Radboud University, 6525 AJ Nijmegen, The Netherlands}
\email{a.levan@astro.ru.nl}

\author[0000-0002-4571-2306]{Jens Hjorth}
\affiliation{DARK, Niels Bohr Institute, University of Copenhagen, Jagtvej 155A, 2200
Copenhagen N, Denmark}
\email{jens@nbi.ku.dk}

\correspondingauthor{Xiao-Fei Dong}
\email{x.dong.2@research.gla.ac.uk}

%%%%%%%%%%%%%%%%%%%%%%%%%%%%
%%========Abstract=======%%%
%%%%%%%%%%%%%%%%%%%%%%%%%%%%
\begin{abstract}
Kilonovae provide a key electromagnetic window into binary neutron star mergers, revealing the properties of the merging system while probing r-process nucleosynthesis of the Universe. However, only a few kilonova candidates have been detected to date, with AT2017gfo being the only one associated with a gravitational wave event, GW170817. In this work, we present \genova, a probabilistic framework for predicting kilonova spectra and light curves directly from gravitational wave posterior samples of binary neutron star mergers. This method uses a conditional normalising flow to learn the distribution of rest-frame spectra conditioned on the source-frame component masses, tidal deformabilities, viewing angle and time since merger. Other kilonova model parameters, such as ejecta opacities and lanthanide opening angle, are marginalised over during training, so that their effects are propagated into the predicted spectra as model predictive uncertainty. In the self-consistency test against the original kilonova model, the flow model reproduces the median light curves with residuals typically below $\sim 0.1$ mag, and the ratio of the predicted central 68\% interval widths remains predominantly between $0.8$ and $1.4$ over $\sim 0.4$--$8.0$ days post merger. Comparisons with a physically distinct kilonova model further show that the probabilistic prediction from the normalising flow can remain informative beyond the model used for training. We apply \genova to GW170817/AT2017gfo using multi-band observations, including newly re-reduced $Y$-, $J$-, $K_s$-band photometry from the Visible and Infrared Survey Telescope for Astronomy, which we present in this work. The resulting predictive intervals broadly encompass the observations while capturing both gravitational wave posterior uncertainty and the variation induced by marginalised kilonova model parameters. These results establish a proof of concept for an efficient probabilistic framework using a normalising flow for predicting kilonova observables directly from gravitational waves.
\end{abstract}

\keywords{\uat{High Energy astrophysics}{739}, \uat{Gravitational wave astronomy}{675}, \uat{light curves}{918}}

%%%%%%%%%%%%%%%%%%%%%%%%%%%%%%%%
%%========Introduction=======%%%
%%%%%%%%%%%%%%%%%%%%%%%%%%%%%%%%
\section{Introduction} \label{sec:intro}
\setcounter{footnote}{0}
Kilonovae (KNe) are the optical/near-infrared (NIR) counterparts to gravitational wave (GW) emission, arising from the radioactive decay of heavy elements formed through rapid neutron captures (r-process) \citep{thi11_rprocess} that can follow binary neutron star (BNS) mergers \citep{li1998}. Before the first direct GW detection of a BNS merger, observational evidence for KNe was primarily sought in association with short gamma-ray bursts (GRBs). The NIR excess observed in GRB~130603B provided the first strong KN candidate associated with a short GRB \citep{Tan13GRB}, with further KN candidates later identified in other GRB events \citep[e.g.,][]{Jin16,Yang15,Lamb19,Tro19}. These observations established an early GRB--KN connection, suggesting that at least some short GRBs originate from the same compact binary mergers that can produce KN emission.

This picture was directly confirmed by the landmark multimessenger observations of GW170817 \citep{abb17a,abb17b}, a BNS merger detected through GWs and accompanied by the short GRB~170817A \citep{abb17GRB,Gol17GRB,Sav17GRB}, the KN AT2017gfo \citep[e.g.,][]{cou17,soa17,val17,arc17,Cow17,dro17,Eva17UV170817,pian17,sma17,kasl17,tan17,hu17,diaz17,Lip17,And17,uts17,sha17,cho17,Nicholl17,mcc17,kil17,poz18}, and a multi-wavelength afterglow \citep[e.g.,][]{tro17,hag17,hall17,ale17,Marg17,Marg18,lym18}. This event marked a major milestone in GW-led multi-messenger astronomy, providing the first compelling evidence that GWs, short GRBs, and KNe can originate from the same NS merger. The combined GW--EM observations enabled constraints on the equation of state (EOS) of neutron-rich dense matter \citep[e.g,][]{mar17,rad18EOS,die20,bres21,bres24}, provided an observational channel for studying heavy $r$-process nucleosynthesis in the Universe \citep[e.g.,][]{met10,kas17,dro17,sma17,pian17}, and allowed for independent measurements of the Hubble constant $H_{\rm 0}$ (see \citealt{bu22} for a review). 

The observations of AT2017gfo also played a central role in shaping current kilonova modelling. Earlier analytic and radiative-transfer studies had already shown that the emission is determined by ejecta mass, velocity, radioactive heating, composition, etc \citep[e.g.][]{li1998,met10,bar13,kas13,tanaka13}. However, the rapid transition of AT2017gfo from blue optical emission to redder optical/NIR emission provided strong motivation for multi-component models, with different ejecta components contributing on different timescales and wavelengths \citep[e.g.][]{Cow17,dro17,sma17,kas17,Per17,tan17,vil17,nic21}. In this picture, the early blue emission is associated with relatively lanthanide-poor and hence low opacity material, while the later redder emission is produced by lanthanide-rich dynamical ejecta and/or disk-wind outflows. In addition, ejecta geometry, viewing angle effects, are important in kilonova modelling \citep[e.g.][]{bu23}.

Despite the importance of joint GW-KN observations, GW170817 remains the only confirmed event with both a GW detection and a well-observed kilonova counterpart. With the improving sensitivity of current and future GW detectors, more BNS merger candidates are expected to be detected. One of the important tasks is therefore how to use the information available from GW observations to rapidly predict the possible KN emission and hence inform EM follow-up observations.

Predicting kilonova emission from GW observables requires connecting initial binary constraints to ejecta properties and a further detailed radiative transfer modelling. GW parameter estimation provides posterior samples for source parameters such as the component masses $(m_1,m_2)$, dimensionless tidal deformabilities $(\Lambda_1,\Lambda_2)$, and the inclination angle $\theta_{\mathrm{JN}}$. These quantities provide physically motivated inputs for kilonova modelling, as the binary parameters directly affect key properties of the dynamical ejecta, such as its mass and characteristic velocity \citep[e.g.,][]{Die17,cou19,die20,ned22,kru20}, while the inclination angle sets the viewing orientation. 

The source parameters do not uniquely determine the kilonova spectra or light curves. The final kilonova observables also depend on further assumptions about both the dynamic and post-merger ejecta, and radiative transfer physics. These include the ejecta geometry and viewing angle dependence \citep[e.g.,][]{Per17,bu19,wol21}, the ejecta composition and opacity \citep{kas13,kas17,tanaka13,bu23,ban20,ban22,ban24}, the radioactive heating rate and thermalisation efficiency \citep{bu23,ros24hr}, the disk-wind contribution \citep[e.g.,][]{com23}, and possible additional energy sources \citep[e.g.,][]{nic21,ai25}, which are either not directly or only weakly constrained by the GW signal. Therefore, their treatment depends on the adopted kilonova model and hence introduces an additional source of astrophysical uncertainty. 

The uncertainty from propagating GW posterior samples to electromagnetic observables involves evaluating the kilonova forward model over both the GW-inferred binary parameters and the remaining model-dependent quantities. For large posterior ensembles, scalability therefore becomes an important practical consideration, especially when the aim is to construct predictive distributions of spectra and light curves rather than a single deterministic prediction. In this sense, surrogate models can provide an efficient approximation to the ensemble forward calculation while retaining the uncertainty associated with model-dependent kilonova parameters.

Recently, machine learning surrogates and simulation-based inference methods have increasingly been explored to accelerate kilonova modelling and inference, including spectra and light curve emulators \citep[e.g.][]{luk22,saha24,ris22,ked23,peng24}, likelihood-free approaches for rapid KN parameter estimation \citep[e.g.][]{darc24,des25,bro26}, and broader multi-messenger inference frameworks for kilonova, GW, and GRB-afterglow observations \citep[e.g.][]{pang23,koe25}. Much of this work targets rapid inference within ejecta parameterisations from EM data. In this work, we focus on a complementary GW-led forward problem: predicting the range of kilonova spectra and light curves implied by GW-constrained binary parameters. To achieve this, we develop a proof of concept probabilistic surrogate framework, \genova, using a conditional normalising flow, which maps GW posterior samples to distributions of kilonova spectra while marginalising over the less constrained kilonova model parameters. This allows both the GW inferred binary uncertainty and the model-dependent kilonova uncertainty to be propagated into the predicted light curves, while establishing a foundation for incorporating richer
physical descriptions and multiple kilonova prescriptions in future applications.

We demonstrate the framework by training and validating the surrogate within the adopted kilonova model, testing its predictions against an independent radiative-transfer grid, and applying it to GW170817/AT2017gfo. The paper is structured as follows: Section \ref{sec:method} describes the kilonova model, the normalising flow framework, and the data preparation; Section \ref{sec:results} presents the surrogate validation, cross-model tests, and application on GW170817/AT2017gfo; In Section \ref{sec:discussion}, we discuss the implications and limitations obtained from the test and application of \genova; We finally summarises our main conclusions in Section \ref{sec:conclusion}.

%%%%%%%%%%%%%%%%%%%%%%%%%%%%%%%
%%========Methodology=======%%%
%%%%%%%%%%%%%%%%%%%%%%%%%%%%%%%

\section{Methodology} \label{sec:method}

\subsection{Kilonova modelling}
This work focuses on the three-component kilonova model introduced by \citet{nic21} (hereafter referred to as \nichollbns), using its implementation in \redback\footnote{Within \redback, this model is available under the model name \bns; the implementation can be found at \url{https://github.com/nikhil-sarin/redback/}.} \citep{sar24a}. In this work, the model generates kilonova spectra and light curves from the initial binary neutron star properties, namely component masses $(m_1,m_2)$ and dimensionless tidal deformabilities $(\Lambda_1, \Lambda_2)$, and additional parameters accounting for the ejecta properties and viewing angle effect. Here, instead of using the original radioactive heating rate based on simulations for an electron fraction $Y_e\simeq 0.04$, we adopt the updated raw heating rate prescription \citep{ros24hr,sar24b}:
\begin{equation}\label{eq:hr_new}
\begin{aligned}
\dot{q}(t)={}&q_0
\left(
 \frac{1}{2} - \frac{1}{\pi}
   \arctan\!\left(\frac{t-t_0}{\sigma}\right)
\right)^{\alpha}
\\
&\times
\left(
 \frac{1}{2} + \frac{1}{\pi}
   \arctan\!\left(\frac{t-t_1}{\sigma_1}\right)
\right)^{\alpha_1}
\\
&\quad
+ C_1 e^{-t/\tau_1}
+ C_2 e^{-t/\tau_2}
+ C_3 e^{-t/\tau_3}. 
\end{aligned}
\end{equation}
where $t$ is the rest-frame time in days, which is defined such that $t=0$ at merger\footnote{Throughout this work, $t$ represents the rest-frame time since the binary merger. In the adopted kilonova modelling, this is also used as the ejecta expansion time in the rest frame, therefore neglecting the time delays between the merger and mass ejection.}, $q_0$, $\alpha$, $t_0$, $\sigma$, $\alpha_1$, $t_1$, $\sigma_1$, $C_i$, and $\tau_i$ ($i=1,2,3$) are fitting coefficients that depend on the velocity and electron fraction of the ejecta. In the equation, the first factor models a constant heating rate with $t_0$ representing the time where there is a transition to power-law decay, while $q_0$ set the normalisation, $\alpha$ is the power-law index, and $\sigma$ is the transition width. The second factor describes the subsecond rise of the heating rate, where $\alpha_1$, $t_1$, and $\sigma_1$ denote the corresponding power-law index, threshold time, and transition width, respectively. The final three terms are included for cases in which individual radioactive isotopes dominate the heating rate; $C_i$ and $\tau_i$ denote their coefficients and decay timescales. We refer to the Tables~A1 and A2 of \citet{ros24hr} for the specific fitting coefficients. For the numerical implementation adopted here, we follow footnote~5 of \citet{sar24b} using \redback. All other model ingredients follow the original \nichollbns prescription of \citet{nic21}. This model then serves as a practical forward generator for constructing a large synthetic spectra dataset that can be used for machine learning model development, while keeping a physically motivated connection between the binary parameters and the resulting kilonova emission. 

In this work, our main focus are the initial binary parameters that can be shared with both GW parameter estimation and kilonova simulation, namely $\theta \sim (m_1,m_2,\Lambda_1,\Lambda_2,\mu_{\mathrm{obs}})$, where $(m_1,m_2,\Lambda_1,\Lambda_2)$ represent the source-frame component masses and dimensionless tidal deformabilities, and $\mu_{\mathrm{obs}}=\cos\theta_{\mathrm{obs}}$ accounts for the viewing angle effect when observing the kilonova. For applications to GW events, we take the geometry information directly from the posterior samples of $\theta_{\rm JN}$ in GW parameter estimation, which is defined as the angle between the line of sight and the total angular momentum of the binary system. Here, we assume that the ejecta symmetry axis is aligned with the binary total angular momentum and restrict $0\leq\theta_{\rm obs}\leq\pi/2$ using the symmetry of the model with respect to the two polar directions. Under this mapping, $\mu_{\mathrm{obs}}\simeq |\cos\theta_{\mathrm{JN}}|$, and $\mu_{\rm obs}$ is used as the viewing angle input throughout this work. Other additional parameters $\boldsymbol{\eta} = (M_{\mathrm{TOV}}, \kappa_{\mathrm{blue}}, \kappa_{\mathrm{red}}, T_{\mathrm{floor,1-3}}, \alpha, \epsilon, \mu_{\mathrm{open}},\mu_{\mathrm{cocoon}})$, which are not directly constrained by GW observation but required for generating the kilonova observables in this model, are considered as nuisance parameters in the normalising flow framework described in Section~\ref{subsec:flow}. The definitions of all adopted parameters are listed in Table~\ref{tab:bns_para}. 

\begin{figure*}[htbp]
\centering
    \includegraphics[width=0.8\textwidth]{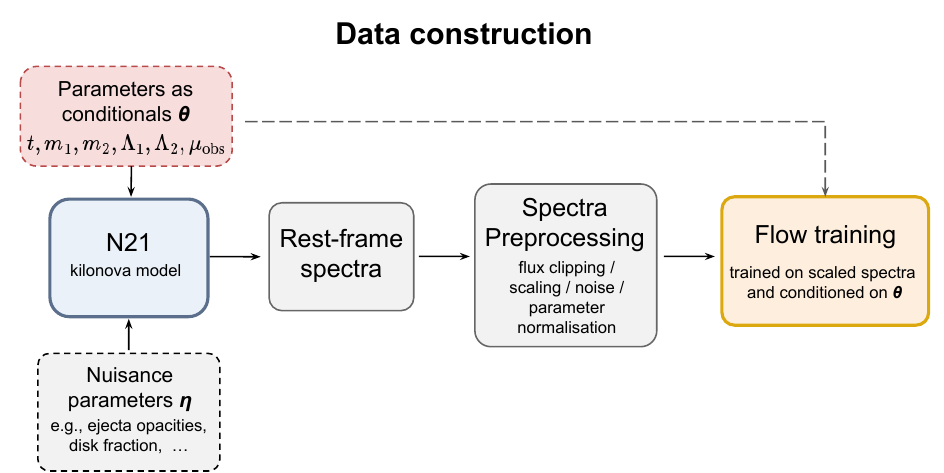}
    \caption{Schematic overview of the training data construction for \genova. In the training stage, binary parameters and additional kilonova nuisance parameters are sampled within the adopted parameter ranges with astrophysical constraints. See Table \ref{tab:bns_para} for the adopted parameter definitions and initial sampling distributions. They are then passed to the \nichollbns{} model to generate source frame spectra, which are preprocessed through flux clipping and scaling, noise treatment, and parameter normalisation before being used to train the model.}
    \label{fig:data_construction}
\end{figure*} 

\subsection{A normalising flow probabilistic framework} \label{subsec:flow}

\begin{figure*}[htbp]
\centering
    \includegraphics[width=0.8\textwidth]{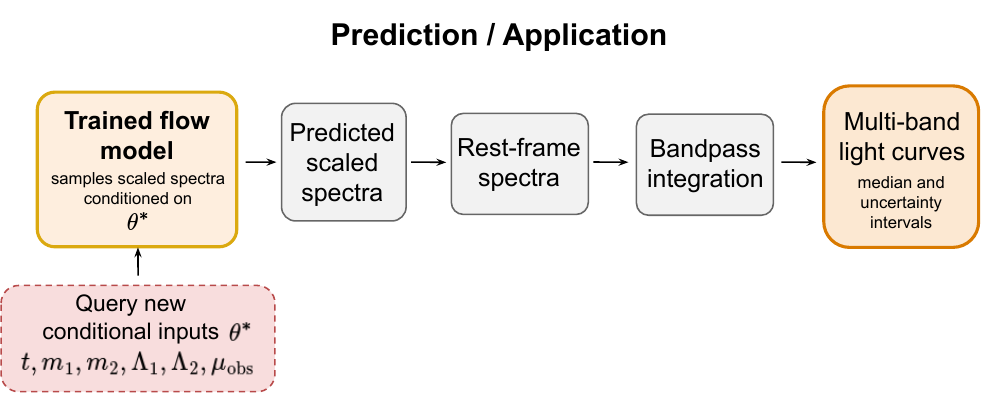}
    \caption{Schematic overview of the prediction processes of \genova. In the prediction stage, the trained flow takes a new query $\boldsymbol{\theta^*}$ and predicts scaled spectra, which are then returned to flux density units by applying the inverse scaling. Given the luminosity distance and redshift information, the spectra are further converted to the observer frame and integrated through photometric bandpasses to generate final multi-band light curves with corresponding uncertainty intervals.}
    \label{fig:pred_flow}
\end{figure*} 

Normalising flows represent a class of generative machine learning models designed to learn complex probability distributions through a series of invertible transformations. These models map an implicit data distribution into a simple latent distribution (e.g., a standard Gaussian) using a series of bijective functions \citep{kob21}. During the training process, an input sample $x \in \mathcal{X}$ is guided by provided variables $\theta \in \Theta$, where $\mathcal{X}$ and $\Theta$ denote the data space and the conditional space, respectively. The model learns a series of explicit transformation functions~$f$ that map the data distribution to a base distribution in the latent space $\mathcal{Z}$, with the latent variable denoted as $z = f(x; \theta)$. The probability density of the conditional distribution is evaluated using the change-of-variables formula,
\begin{equation}
p_{\mathcal{X}|\Theta}(x \mid \theta)
=
p_{\mathcal{Z}}\!\left(f(x;\theta)\right)
\left|\det J_f(x;\theta)\right|,
\end{equation}
where $J_f(x;\theta)=~\partial f(x;\theta)/\partial x$ is the Jacobian of the transformation with respect to the data variable $x$. Once trained, the model allows for efficient predictions by sampling from the latent space and applying the inverse functions $f^{-1}$. By learning the conditional distribution ($p(x\mid \theta)$) explicitly, the normalising flow provides an efficient surrogate for the original numerical/analytical models, while preserving the corresponding probabilities.

In this work, we develop a conditional normalising flow to predict kilonova spectra from GW-derived binary parameters for binary neutron star mergers, with other KN parameters marginalised. In our setup, the data space consists of theoretical spectra $\mathbf{x}$ evaluated on a fixed wavelength grid with $20$ points $\mathbf{x}\in\mathbb{R}^{20}$, while the conditional space contains the associated physical parameter set $\boldsymbol{\theta}= (t, m_1, m_2, \Lambda_1, \Lambda_2, \mu_{\mathrm{obs}})$ on which each spectrum is conditioned. The latent space provides a twenty-dimensional standard Gaussian representation that makes sampling straightforward.

Specifically, as illustrated in Figure~\ref{fig:data_construction}, we pre-generate the dataset to perform the training, validation, and testing for the flow model:
\begin{equation}
\mathcal{D} = \{(\mathbf{x}_i, \boldsymbol{\theta}_i, \boldsymbol{\eta}_i)\}_{i=1}^{N},
\end{equation}
where $N$ is the total number of spectra. $\boldsymbol{\eta} = (M_{\mathrm{TOV}}, \kappa_{\mathrm{blue}}, \kappa_{\mathrm{red}}, T_{\mathrm{floor,1-3}}, \alpha, \epsilon, \mu_{\mathrm{open}},\mu_{\mathrm{cocoon}})$ collects the remaining simulation parameters that vary across the dataset but are not explicitly provided to the flow model. Therefore, the flow predicts the conditional spectral distribution in the projected $(\mathbf{x}, \boldsymbol{\theta})$ space, rather than in the joint space of $(\mathbf{x}, \boldsymbol{\theta}, \boldsymbol{\eta})$. In this sense, the effect of the unconditioned parameters $\boldsymbol{\eta}$ is absorbed implicitly into the learned distribution, with the corresponding probability written as
\begin{equation}
p_{\phi}(\mathbf{x}\mid\boldsymbol{\theta})
\approx
p_{\rm train}(\mathbf{x}\mid\boldsymbol{\theta})
=
\frac{
\int p_{\rm train}(\mathbf{x},\boldsymbol{\theta},\boldsymbol{\eta})
\,{\rm d}\boldsymbol{\eta}
}{
p_{\rm train}(\boldsymbol{\theta})
},
\label{eq:marginal1}
\end{equation}
where $\phi$ denotes the trainable parameters of the flow model, and $p_{\rm train}$ is the probability distribution represented by the training dataset, with different arguments indicating the corresponding joint, marginal, or conditional distribution. Once the flow model is trained, we can easily draw samples from the latent space for any conditional input $\boldsymbol{\theta}=(t,\boldsymbol{\vartheta})$ within the support of the training distribution, where $t$ is time and $\boldsymbol{\vartheta}$ contains the remaining physical conditional parameters. For a fixed $\boldsymbol{\theta}$, enough repeated latent draws sample the predictive distribution $p_{\phi}(\mathbf{x}\mid\boldsymbol{\theta})$. More generally, when the remaining conditional parameters themselves form a distribution $q(\boldsymbol{\vartheta})$, the corresponding predictive distribution at a given time $t$ can be obtained by marginalising over these parameters as
\begin{equation}
\begin{aligned}
p_{\phi,\widetilde{q}}(\mathbf{x} \mid t)
&=
\int
p_{\phi}(\mathbf{x}\mid t, \boldsymbol{\vartheta})
\widetilde{q}(\boldsymbol{\vartheta})
\,{\rm d}\boldsymbol{\vartheta} 
\\
&\approx
\frac{1}{N_{\vartheta}}
\sum_{i=1}^{N_{\vartheta}}
p_{\phi}
\left(
\mathbf{x}\mid t, \widetilde{\boldsymbol{\vartheta}}_i
\right),
\quad
\widetilde{\boldsymbol{\vartheta}}_i
\sim
\widetilde q(\boldsymbol{\vartheta}),
\end{aligned}
\label{eq:marginal2}
\end{equation}
where $\widetilde{q}(\boldsymbol{\vartheta})$ represents $q(\boldsymbol{\vartheta})$ restricted to the support of the flow training conditions, and $N_{\vartheta}$ is the number of conditional parameter sets. Predictions from this marginal distribution can be produced by drawing one or more latent samples for each $\widetilde{\boldsymbol{\vartheta}}_i$. When a sufficiently number of conditional inputs is available, one single independent draw for each $\widetilde{\boldsymbol{\vartheta}}_i$ already provides a Monte Carlo sample from the predicted distribution marginalised over both nuisance parameters in Eq.~\ref{eq:marginal1} and the distribution of conditional inputs in Eq.~\ref{eq:marginal2}. The resulting ensemble predictions therefore represents a probabilistic prediction of spectra for the given conditional inputs distribution. 
Since all spectra in this work are generated in the rest frame at a fiducial distance of $10$~pc, we can obtain the absolute AB magnitudes $M_{b}$ for a given bandpass $b$, which is described as
\begin{equation}
M_{b}(t)=-2.5\log_{10}
\left[
\frac{
\displaystyle \int \lambda S_b(\lambda)\,
\widehat{F}_{\lambda}^{10{\rm pc}}(t,\lambda)\,d\lambda
}{
\displaystyle \int \lambda S_b(\lambda)\,
F_{\lambda,0}^{\rm AB}(\lambda)\,d\lambda
}
\right],
\label{eq:abmag}
\end{equation}
where $S_b(\lambda)$ is the throughput of band $b$, $F_\lambda$ denotes the spectral flux density per unit wavelength. Since the predicted spectra are available only at discrete wavelength points, $\widehat{F}_\lambda$ specifically denotes the interpolated representation constructed from the predicted flux densities. The quantity $F_{\lambda,0}^{\rm AB}$ is the AB reference flux density in wavelength units, corresponding to $F_{\nu,0}=3631~{\rm Jy}$. 

When luminosity distance $d_{\mathrm{L}}$ is available, the corresponding redshift $z$ can be obtained by assuming a specific cosmological model. The $10$~pc rest-frame spectrum can then be converted accordingly to the observer frame as
\begin{equation}
\widehat{F}_{\lambda}^{\rm obs}(t_{\rm obs},\lambda_{\rm obs})
=
\frac{\left(10~{\rm pc}/d_{\mathrm{L}}\right)^2}{1+z}
\widehat{F}_{\lambda}^{10{\rm pc}}(t_{\rm rest},\lambda_{\rm rest}),
\label{eq:observer_flux_lambda}
\end{equation}
where $t_{\rm obs}=t_{\rm rest} \cdot(1+z)$ and $\lambda_{\rm obs}=\lambda_{\rm rest}\cdot(1+z)$. The apparent AB magnitude is then obtained by replacing $\widehat{F}_{\lambda}^{10{\rm pc}}$ in Eq.~\ref{eq:abmag} with $\widehat{F}_{\lambda}^{\rm obs}$. We use \sncosmo to perform the spectra interpolation and magnitude calculation. This process provides an approximation of the conditional predictive distribution for the magnitude, $p(M_b|\boldsymbol{\theta})$, or $p(m_b|\boldsymbol{\theta}, d_{\mathrm{L}},z)$ when luminosity distance and redshift information are included. Repeating the procedure over a time grid yields probabilistic light curves that propagate both nuisance parameter variation, where applicable, the uncertainty from the conditional inputs. 

Throughout this work, the predicted distributions are characterised by their medians and central $68\%$ and $90\%$ quantile intervals, corresponding to the 50th, 16th--84th and 5th--95th percentiles, respectively. The described prediction workflow is illustrated in Figure~\ref{fig:pred_flow}. To infer redshift from luminosity distance, we adopt the \texttt{Planck18} flat $\Lambda$CDM cosmology implemented in \astropy, with $H_0 = 67.66~{\rm km~s^{-1}~Mpc^{-1}}$ and $\Omega_{\rm m}=0.31$ \citep{planck2018}. 

\subsection{Data preparation and training setup} 
\label{subsec:datapre}

\begin{figure}[htbp]
\centering
    \includegraphics[width=\columnwidth]{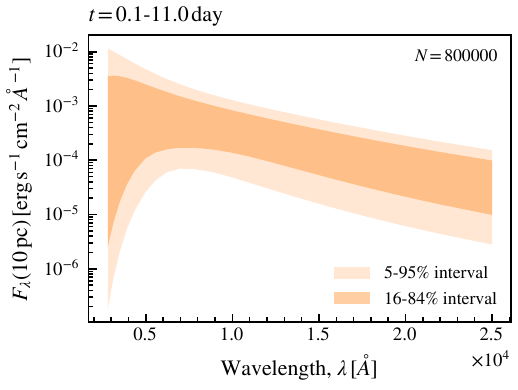}
    \caption{Distribution of training spectra before rescaling. The orange ranges show the central 68\% and 90\% intervals of the training set within the training time range, $0.1$--$11.0$ days post merger. The number of training samples is marked on the top right corner.}
    \label{fig:spec_samples}
\end{figure}

To construct the spectra dataset mentioned in Section~\ref{subsec:flow}, we first sample $(m_1,m_2,\Lambda_1,\Lambda_2)$ combinations following the initial distributions listed in the upper block of Table~\ref{tab:bns_para}. During the generation, we keep $m_1\geq m_2$, and only combinations satisfying the binary consistency criteria are retained to avoid extreme combinations that are less likely astrophysical. These criteria include $\Lambda_1\leq\Lambda_2$, the allowed ranges and ordering relations for the component radii $R_{1,2}$ and compactnesses $\mathcal{C}_{1,2}$, and the range of the mass ratio $q=m_2/m_1$. All criteria are listed in the middle block in Table~\ref{tab:bns_para}. We finally obtain $10^6$ distinct combinations of $(m_1,m_2,\Lambda_1,\Lambda_2)$ in total though this process. For each combination, we independently draw one set of nuisance kilonova parameters from the distributions listed in the lower block of Table~\ref{tab:bns_para} to generate spectra. This results in a total of $10^6$ spectra at all times in the entire dataset. Each spectrum is computed from $2800-25000$\ang with $20$ log-spaced wavelength bins at a distance of $10$~pc in the rest frame. They are then clipped below at $10^{-8}$~\flux. Figure~\ref{fig:spec_samples} shows the central $68\%$ and $90\%$ quantile ranges of the samples used for training after clipping. We further rescale the spectra by a factor of $100$ to bring their typical values closer to unity. We then add zero-mean Gaussian noise with a standard deviation of $10^{-3}$ to improve the convergence of the flow training. Each conditional parameter is independently normalised to $[0,1]$ using its initial sampling bounds to place the inputs on comparable numerical scales before training. During prediction, the spectra are transformed back to physical units by reversing the preprocessing scaling. 

\begin{deluxetable*}{lll}
\tabletypesize{\small}
\tablecaption{
Definitions, initial sampling distributions, and binary consistency
criteria adopted in constructing the \nichollbns kilonova
training set.
\label{tab:bns_para}}
\tablewidth{0pt}
\tablehead{
\colhead{Parameter}
&
\colhead{Definition}
&
\colhead{Distribution/Criterion}
}
\startdata
\multicolumn{3}{c}{\textit{Binary input parameters}} \\
\tableline
$t/\mathrm{days}$
& Rest-frame time since merger
& $\operatorname{log}\mathcal{U}(0.1,\,11.0)$ \\
$m_{1,2}/M_\odot$
& Source-frame component masses
& $\mathcal{U}(1.1,\,1.8)$ \\
$\Lambda_{1,2}$
& Dimensionless component tidal deformabilities
& $\mathcal{U}(10,\,3000)$ \\
$\mu_{\rm obs}$
& Cosine of the viewing angle $\theta_{\rm obs}$
& $\mathcal{U}(0,\,1)$ \\
\tableline
\multicolumn{3}{c}{\textit{Binary consistency criteria}} \\
\tableline
$q$
& Mass ratio, $q=m_2/m_1$
& $q\in[0.6,\,1.0]$ \\
$\Lambda_{1,2}$
& Tidal deformability ordering
& $\Lambda_1\leq\Lambda_2$ \\
$\mathcal{C}_{1,2}$
& Component compactnesses
& $\mathcal{C}_{1,2}\in[0.1,\,0.23],\quad
   \mathcal{C}_1\geq\mathcal{C}_2$ \\
$R_{1,2}/\mathrm{km}$
& Component radii
& $R_{1,2}\in[9,\,15],\quad R_1\leq R_2$ \\
\tableline
\multicolumn{3}{c}{\textit{Nuisance kilonova parameters}} \\
\tableline
$M_{\rm TOV}/M_\odot$
& Tolman--Oppenheimer--Volkoff mass
& $\mathcal{U}(2.05,\,2.4)$ \\
$\kappa_{\rm blue}/(\mathrm{cm}^{2}\,\mathrm{g}^{-1})$
& Gray opacity of the blue ejecta component
& $\mathcal{U}(0.01,\,1.0)$ \\
$\kappa_{\rm red}/(\mathrm{cm}^{2}\,\mathrm{g}^{-1})$
& Gray opacity of the red ejecta component
& $\mathcal{U}(1.0,\,30.0)$ \\
$T_{{\rm floor},i}/\mathrm{K}$
& Temperature floor of ejecta component $i=1,2,3$
& $\mathcal{U}(2000,\,4000)$ \\
$\alpha$
& Enhancement of blue ejecta by neutron star surface winds
& $\mathcal{U}(0.1,\,1.0)$ \\
$\epsilon$
& Fraction of the disk mass that becomes unbound
& $\mathcal{U}(0.1,\,0.3)$ \\
$\mu_{\rm open}$
& Cosine of the lanthanide opening angle
& $\mathcal{U}(0.5,\,0.866)$ \\
$\mu_{\rm cocoon}$
& Cosine of the shocked-cocoon opening angle
& $\mathcal{U}(0.866,\,1.0)$ \\
\enddata
\tablecomments{
The distributions in the upper and lower blocks describe only the initial parameter sampling distributions. The binary components are labelled such that $m_1\geq m_2$, and sampled combinations are retained only when all criteria in the middle block are satisfied. The mass ratio $q$, component compactnesses $\mathcal{C}_{1,2}$, and radii $R_{1,2}$ are derived from the sampled binary parameters. The compactnesses are calculated using the $\Lambda$--$\mathcal{C}$ quasi-universal relation adopted by \citet{nic21}.
}
\end{deluxetable*}
The final dataset is split into training, validation, and test sets in the $(m_1,m_2,\Lambda_1,\Lambda_2)$ space, using 80\% of the samples for training, 10\% for validation, and 10\% for test. The validation set is held out from the optimisation process and is used to monitor the generalisation performance of the normalising flow model. In this work, we use a real-valued non-volume preserving (RealNVP) normalising flow \citep{din17}, constructed from stacked coupling transforms and implemented with \gla \citep{mic24}, which builds on \nflows \citep{con20} and \pyt \citep{pas19}. The model is trained by minimising the Kullback--Leibler (KL) divergence \citep{pap19} from the empirical training distribution to the learned conditional distribution. The adopted model architecture and training configuration are described in Appendix~\ref{app:flow_config}.

\begin{figure}[ht]
\centering
    \includegraphics[width=\columnwidth]{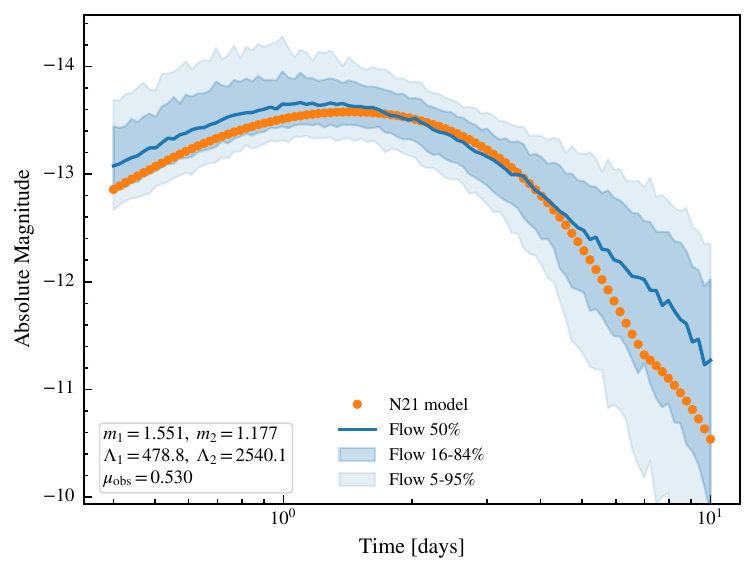}
    \caption{Example light curve comparison for a test case. The corresponding $(m_1,m_2,\Lambda_1,\Lambda_2)$ values are drawn from the test set. The predicted $r$-band light curves from the flow model are shown by the median and central 68\% quantile interval, compared with the theoretical \nichollbns light curve. Note that the median curve should be interpreted as a summary of the marginal predicted distribution, rather than a reconstruction of the individual simulated example shown for comparison.}
    \label{fig:testlc}
\end{figure}

%%%%%%%%%%%%%%%%%%%%%%%%%%%
%%========Results=======%%%
%%%%%%%%%%%%%%%%%%%%%%%%%%%
\section{Results} \label{sec:results}
In this section, we evaluate the trained flow model using the prediction workflow described in Section~\ref{subsec:flow}. We first compare \genova predictions with the corresponding \nichollbns outputs. We then perform a cross-model comparison with the radiative transfer models chosen from \citet{kas17}\footnote{The models are available at \url{https://github.com/dnkasen/Kasen_Kilonova_Models_2017}.} (hereafter referred to as \kasen). Finally, we apply this workflow to the real detections of GW170817/AT2017gfo. 

\subsection{In-distribution tests}
We first inspect representative in-distribution examples to illustrate the form of the conditional predictive distribution, and then quantify the agreement with reference \nichollbns light curves over the test set.

We begin by showing a representative in-distribution example for a set of test conditional inputs. The reference \nichollbns light curves correspond to a fixed set of nuisance parameters, and are therefore used here as qualitative consistency checks.

Figure~\ref{fig:testlc} shows the flow predictions in $r$-band for a selected example in the test sets. For each binary conditional input, the flow prediction is summarised by the median and central 68\% and 90\% quantile intervals, while the reference light curve is generated from the \nichollbns model using the same binary parameters with a fixed set of nuisance parameters taken within the model's supporting range. The flow prediction is evaluated at $100$ logarithmically spaced time points from $0.4$ to $10$ days after merger. At each time, we draw $4000$ latent samples, map them back to the spectra space, and convert the predicted spectra into magnitudes as described to construct the corresponding light curves. Full comparison plots in other representative photometric bands are shown in Figure~\ref{fig:test_bands} in Appendix~\ref{app:test_plots}. Overall, the reference light curve is consistent with the predicted distribution for the test example. The reference light curve is generated using a particular draw of the nuisance parameters and hence represents one possible realisation of the nuisance-marginalised distribution, whereas the shaded region summarises the corresponding distribution predicted by the flow. The purpose of this comparison is to access whether the predicted quantile intervals encompass the example \nichollbns light curve. The consistency between the reference light curve and the blue interval demonstrates that the flow captures the conditional training distribution and generalises to unseen inputs.

While the test example illustrates the general agreement between the flow predictions and the theoretical light curves, a statistical summary over the full testing set is needed to assess the model behaviour more systematically. We evaluate the flow model on the test set containing $10^5$ cases. Each test case $i$ corresponds to a complete set of multi-band light curves evaluated on a common time grid $\{t_k\}$, consisting of $20$ logarithmically spaced points over the time range of $[0.4,10]$ days after merger. Specifically, for each test case $i$ and time point $t_k$, we draw $N_{\mathrm{draw}}=500$ samples from the latent space using the trained flow model to form the predictive spectra distribution. These sampled spectra are then integrated over the chosen bandpasses to obtain the corresponding magnitude distributions in each photometric band~$b$. The resulting flow prediction is compared against the corresponding distribution from the \nichollbns model evaluated under the same conditions.

For each test case, time point, and band, we first compare the central behaviour of the two distributions using the residual between their median magnitudes,
\begin{equation}
\Delta M^{50\%}_{i,k,b}
\equiv
M^{\mathrm{flow}}_{50\%,i,k,b}
-
M^{\mathrm{\nichollbns}}_{50\%,i,k,b},
\label{eq:median_residual}
\end{equation}
where $M^{\mathrm{flow}}_{50\%,i,k,b}$ and $M^{\mathrm{\nichollbns}}_{50\%,i,k,b}$ denote the median magnitudes of the flow predictive distribution and the \nichollbns model distribution for the $i$-th test case, time point $t_k$, and band $b$, respectively. For a fixed test case and band, the set of $\Delta M_{50\%,i,k,b}$ over all time points forms a residual light curve. Positive values of $\Delta M_{50\%,i,k,b}$ indicate that the flow prediction is fainter than the \nichollbns model, while negative values indicate that it is brighter.

\begin{figure}[thbp]
\centering
\includegraphics[width=1.0\columnwidth]{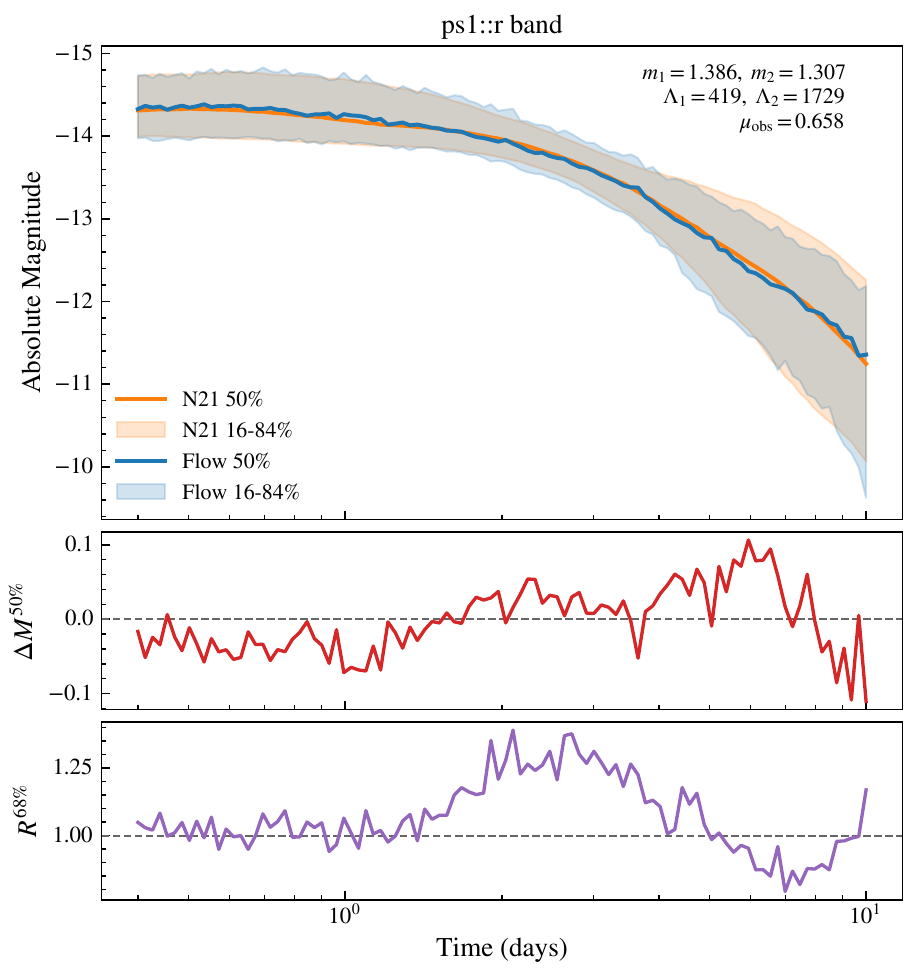}
\caption{Comparison between the \nichollbns model and the flow model for an example test case in the $r$ band. The median and central 68\% quantile interval are shown for both models in the upper panel. The median residuals and central 68\% width ratios are shown in the middle and lower panels, respectively. The $(m_1,m_2,\Lambda_1,\Lambda_2)$ values for this example are drawn from the test set.}
\label{fig:theory_case}
\end{figure}

In addition to the median residual, we compare the widths of the predictive distributions through the ratio of their central 68\% intervals,
\begin{equation}
R^{68\%}_{i,k,b}
\equiv
\frac{
W^{\mathrm{flow}}_{68\%,i,k,b}
}{
W^{\mathrm{\nichollbns}}_{68\%,i,k,b}
},
\label{eq:width_ratio}
\end{equation}
where, for $A \in \{\mathrm{flow}, \mathrm{\nichollbns}\}$, $W^{A}_{68\%,i,k,b} \equiv M^{A}_{84\%,i,k,b} - M^{A}_{16\%,i,k,b}$. $R^{68\%}_{i,k,b}>1$ or $R^{68\%}_{i,k,b}<1$ indicates that the flow predictive distribution is broader or narrower than the theoretical distribution, respectively. Figure~\ref{fig:theory_case} shows an example test case comparison between the predictive distributions from the flow and the \nichollbns model in the $r$~band, where the median residuals and width ratios with time are presented in the bottom two panels. The flow reproduces the time evolution against the reference \nichollbns model, with the median residuals within $\pm 0.1$~mag over evaluated time. The predicted central $68\%$ width is also generally consistent with the reference prediction, with the width ratio differing from unity by up to roughly $20\%-30\%$ at some epochs. As a secondary practical benefit, the flow model also reduces the computational cost of spectra generation. For the example case considered here, using the current implementation, spectra generation with the flow model on a GPU required only $\sim 1/64$ of the runtime of direct generation with \nichollbns. However, we note that this example is intended to illustrate the potential computational benefit instead of providing a systematic performance evaluation. 

\begin{figure}[htbp]
\centering
    \includegraphics[width=1.0\columnwidth]{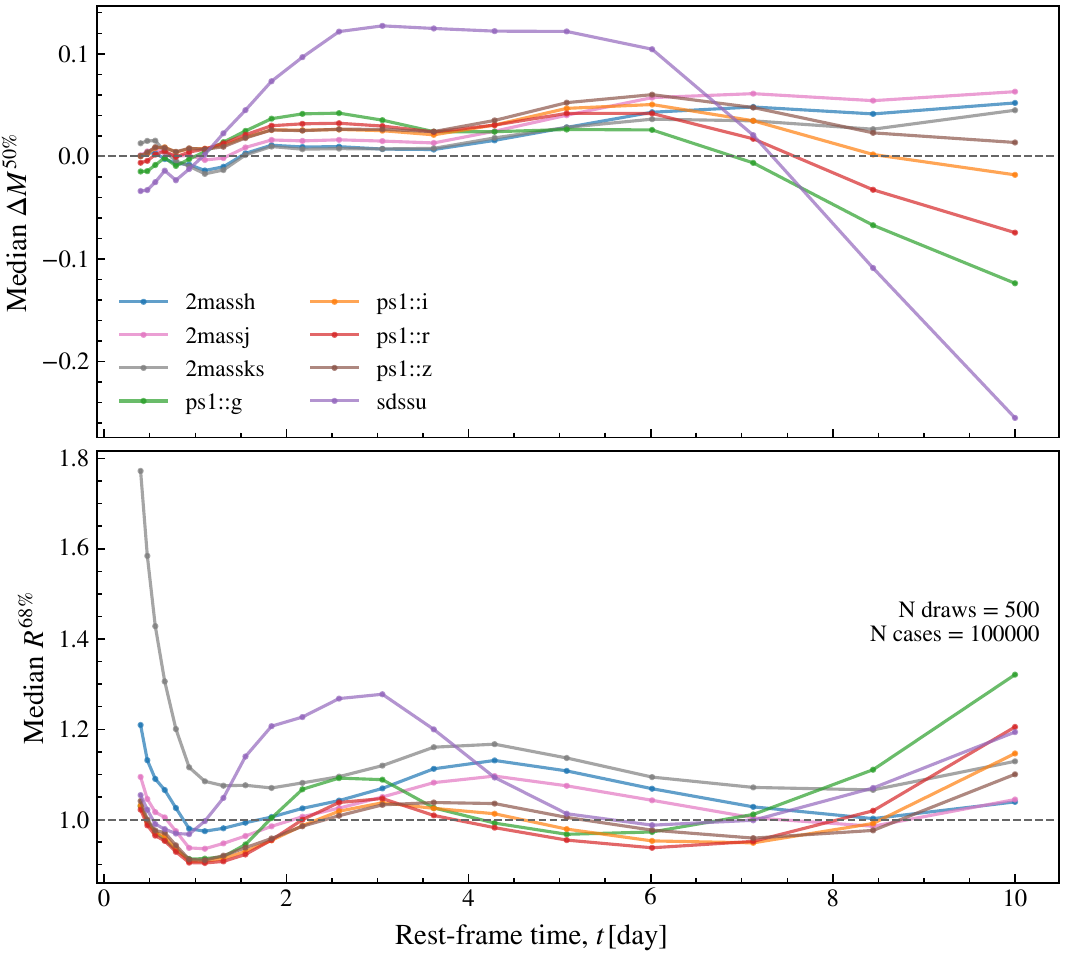}
    \caption{Time evolution of the flow predictive behaviour across $10^5$ test cases in eight bands. Top panel: time-binned median residuals between the flow and the corresponding \nichollbns predictive medians. Bottom panel: time-binned median of the ratio of the central $68\%$ interval widths predicted by the flow and \nichollbns model. Different colours represent different bands.}
    \label{fig:test_error}
\end{figure}

To examine the temporal evolution of the predictive behaviour, we summarise the median residual $\Delta M^{50\%,}$ and the 68\% width ratio $R^{68\%}$ of all test cases as functions of time. These quantities respectively characterise the offset of the flow predictive median from the corresponding \nichollbns model median, and the relative central $68\%$ width of the flow predictive distribution compared to the \nichollbns model distribution. The resulting time-dependent summaries are shown in Figure~\ref{fig:test_error}. In general, the flow model performs well in most bands with the median residual remaining within $\sim 0.1$ mag and the 68\% width ratio are consistently smaller than $\sim 1.4$ between $\sim 0.4-8.0$ days. However, the mismatch for both residual and width ratio at the late time ($\sim 8$ days) is mainly due to the decrease of the brightness, the speed of which is faster from the bluer UV band to redder NIR bands. As the KN becomes dimmer, the flux starts to fall below the noise level, which washes out the true values and hence results in a relatively larger absolute residual and width ratio. Nevertheless, the effect is limited to within $\sim 0.25$~mag for the median residual and $40\%$ for the width ratio. This effect is more pronounced in the $u$~band. As shown in Figure~\ref{fig:spec_samples}, the $u$ band probes the blue part of the spectrum, where the test spectra show a larger dynamical range, especially below $\sim5000$~\ang. The light curves also decline more rapidly after $\sim 1$ day, and can reach $\gtrsim-6$ mags at $10$ day. In this case, as the flux decreases with time, a similar absolute flux error corresponds to a larger fractional error, and is therefore amplified when converted into magnitudes. The residuals in $u$~band therefore most likely reflect the sensitivity of rapidly fading blue fluxes to small flux errors and noise effects at low flux，resulting in a larger median residual of $\sim0.1-0.25$~mag compared to other bands. 

On the other hand, the width ratio in the $K_s$ band reaches $1.2-1.8$ during the first day. This behaviour is likely driven by a combination of spectral resolution and low-flux effects. At early times, the $K_s$ band emission lies on the red, relatively low flux tail of the early spectrum and is close to the red end of the adopted wavelength range. With the current $20$ points log-uniform spectra representation over $2800$--$25000$~\ang, the bandpass is covered by only a small number of wavelength points. The resulting photometry is therefore more sensitive to interpolation uncertainties and to the low-amplitude noise added in the spectral representation. In addition, the reference \nichollbns predictive width is relatively narrow at early times (see Figure \ref{fig:theory_bands} as an example). Consequently, even a modest absolute broadening of the flow prediction can lead to an apparently large width ratio. The same effects are weaker, but still present, in the $J$ and $H$ bands，which probe shorter wavelengths than $K_s$ and are therefore less affected by the sampling at red end and low-flux behaviour.

Nevertheless, the general agreement in both the median evolution and the predictive width over $0.4$--$10$ days supports the use of the flow model for efficient probabilistic light-curve prediction within this time range. The main caveat is that predictions in the bluest ($u$) and reddest ($K_s$) bands require additional care in low-flux regimes, particularly after $\sim 1$ day in the $u$ band and before $\sim 1$ day in the $K_s$ band. In these regimes, interpolation and noise in the learned spectral representation can have an amplified effect on the band-integrated magnitudes. 

\subsection{Cross-model comparison with \kasen light curves}

As an additional test of the predictive behaviour of \genova, we compare the predicted light curve distributions with light curves constructed from \kasen. In contrast to the semi-analytical \nichollbns prescription used to generate the training data, \kasen is parametrised directly by the ejecta properties, including the ejecta mass, velocity, and lanthanide fraction. The models are based on radiative transfer calculations with wavelength-dependent opacities, and therefore provide a physically distinct reference model against which to compare with the \genova predictions.

To perform the comparison, we first select one $(m_1, m_2, \Lambda_1, \Lambda_2)$ combination from the training set of the flow model. We compute the ejecta quantities using the same prescription as adopted in the \nichollbns model, while fixing the remaining kilonova parameters to a reference set.  This gives three ejecta components containing a lanthanide-poor blue dynamical component, a lanthanide-rich red dynamical component, and an intermediate-opacity disk wind component. We extract the ejecta mass and velocity of each component and use these quantities to construct an approximate counterpart in \kasen. \kasen contains single component models parametrised by ejecta mass, ejecta velocity, and lanthanide fraction. Therefore, to compare it with the three-component \nichollbns model, we assign fixed lanthanide fractions of $X_{\rm lan}=10^{-9}$, $10^{-2}$, and $10^{-1}$ to the blue, purple, and red components, respectively, following the values adopted in \citet{nic21}. For each component, we select the nearest \kasen model in ejecta space, and then sum the three component spectra in flux space to obtain the total light curve. Since the \kasen models are not parametrised by viewing angle, we generate the flow predictions at fixed $(t, m_1, m_2, \Lambda_1, \Lambda_2)$ while marginalising over $\mu_{\mathrm{obs}}$. The predicted spectra are then integrated through the same bandpasses and converted to light curves.

\begin{figure*}[t]
\centering
    \includegraphics[width=0.8\textwidth]{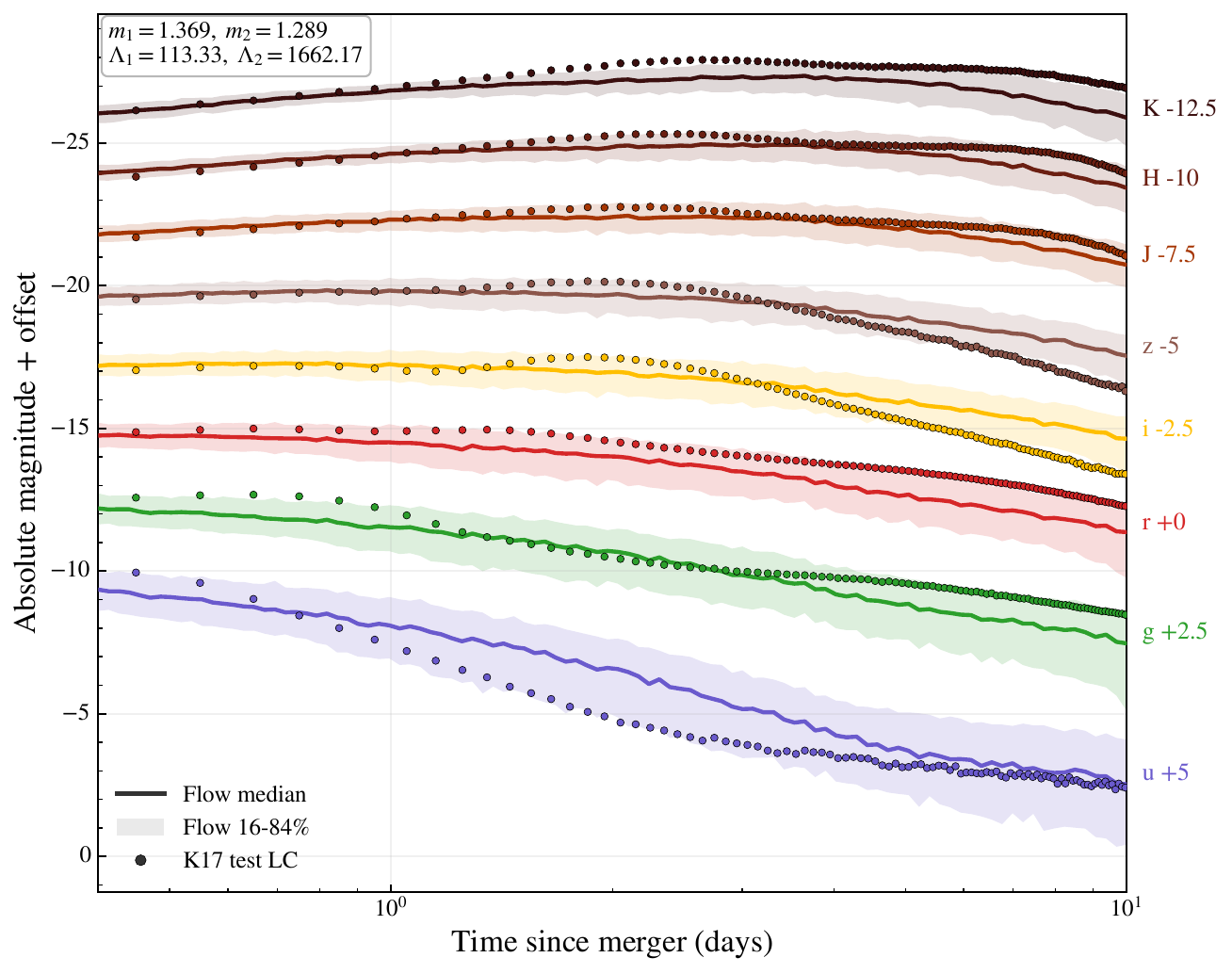}
    \caption{Example cross-model light curve comparison between the flow model predictions and \kasen in the optical/NIR bands. Solid curves show the median flow predictions, while shaded regions indicate the 16--84\% predictive intervals. The points show the corresponding three-component \kasen light curves. Light curves are vertically shifted for clarity, with the applied offsets labelled on the right. The binary parameters used for this example are shown on the upper-left.}
    \label{fig:test_kasen}
\end{figure*}

Figure~\ref{fig:test_kasen} shows an example of this cross-model comparison. Unlike the in-distribution tests, this is not a direct interpolation test within the training parameter space, but a photometric consistency check against light curves constructed from a different radiative transfer model. We emphasise that the purpose of this comparison is not to require the median prediction to reproduce the detailed morphology of the \kasen light curves. Instead, we focus on whether the predicted probabilistic interval provides an informative range for the light curve behaviour. For this selected binary configuration, the \kasen light curves are broadly contained within the central $68\%$ predictive interval of the flow over the $0.4$--$10$ day range in most bands. This suggests that the probabilistic prediction from \genova can remain informative even when applied to light curves generated from a physically distinct reference model. 

\subsection{Application on GW170817}

As a well-detected multi-messenger source in both gravitational wave and electromagnetic regime, GW170817 provides an optimal test ground for the flow model. In this section, we apply \genova to the gravitational wave posterior samples of GW170817, along with the real detections of the corresponding optical counterpart AT2017gfo. We demonstrate the capability of this framework for probabilistic light curve predictions directly using gravitational wave posterior samples.

We take the low-spin posterior samples of GW170817 from GWTC-1 parameter estimation sample release of the LIGO/Virgo Collaboration from \url{https://dcc.ligo.org/LIGO-P1800370/public} \citep{abb19}. 

For the photometric observations of AT2017gfo, we retain the optical $r$-, $i$-, and $z$-band measurements from \citet{tan17}, while for the $Y$, $J$, and $K_s$ bands we use a new reduction of imaging obtained with the European Southern Observatory (ESO) Visible and Infrared Survey Telescope for Astronomy (VISTA; \citealt{vista}). The principal change relative to the previously published VISTA photometry is the use of late-time observations of NGC~4993, obtained on 2019 March 19 after the transient had faded, as reference images for difference imaging. Three 240~s tiled exposures were obtained in each of the $Y$, $J$, and $K_s$ bands. Image subtraction was performed independently for each epoch using the Saccadic Fast Fourier Transform (SFFT) algorithm \citep{SFFT}, with the reference image convolved to match the corresponding science image. This allows the transient flux to be measured after removal of the underlying spatially varying light from NGC~4993. A full description of the image subtraction, calibration, and photometric procedure is given in Appendix~\ref{app:vista_reduction}.

The resulting revised $Y$-, $J$-, and $K_s$-band photometry is presented in Table~\ref{tab:vista_photometry} and replaces the corresponding VISTA measurements reported by \citet{tan17}. The revised values remain closely consistent with the earlier photometry: the residuals have an rms scatter of $0.045$~mag and a mean offset of less than $0.01$~mag. The mean residuals in the individual $Y$, $J$, and $K_s$ bands are $-0.02$, $-0.02$, and $+0.03$~mag, respectively, with all measurements differing by less than $0.1$~mag. All VISTA observations listed here have an exposure time of 120~s. The full revised dataset is reported in the table, while only epochs within the $0.4$--$10$~day prediction range of the flow model are used in the comparison to avoid temporal extrapolation.

\begin{deluxetable}{cccc}
\tablecaption{Revised VISTA near-infrared photometry of AT2017gfo.
\label{tab:vista_photometry}}
\tablewidth{0pt}
\tablehead{
\colhead{$\Delta t$ (d)} &
\colhead{Telescope/Camera} &
\colhead{Filter} &
\colhead{Mag(AB)$_0$} \\
\colhead{(1)} &
\colhead{(2)} &
\colhead{(3)} &
\colhead{(4)}
}
\startdata
0.482 & VISTA/VIRCAM & $Y$   & $17.418 \pm 0.054$ \\
1.467 & VISTA/VIRCAM & $Y$   & $17.185 \pm 0.059$ \\
2.464 & VISTA/VIRCAM & $Y$   & $17.432 \pm 0.056$ \\
3.461 & VISTA/VIRCAM & $Y$   & $17.730 \pm 0.050$ \\
4.459 & VISTA/VIRCAM & $Y$   & $18.039 \pm 0.061$ \\
6.471 & VISTA/VIRCAM & $Y$   & $18.701 \pm 0.086$ \\
7.463 & VISTA/VIRCAM & $Y$   & $19.240 \pm 0.089$ \\
9.457 & VISTA/VIRCAM & $Y$   & $20.167 \pm 0.138$ \\
0.476 & VISTA/VIRCAM & $J$   & $17.781 \pm 0.055$ \\
1.460 & VISTA/VIRCAM & $J$   & $17.419 \pm 0.043$ \\
2.457 & VISTA/VIRCAM & $J$   & $17.634 \pm 0.047$ \\
3.453 & VISTA/VIRCAM & $J$   & $17.828 \pm 0.038$ \\
4.453 & VISTA/VIRCAM & $J$   & $18.058 \pm 0.036$ \\
6.465 & VISTA/VIRCAM & $J$   & $18.719 \pm 0.057$ \\
7.456 & VISTA/VIRCAM & $J$   & $19.069 \pm 0.097$ \\
9.450 & VISTA/VIRCAM & $J$   & $20.071 \pm 0.204$ \\
10.452 & VISTA/VIRCAM & $J$  & $20.967 \pm 0.301$ \\
11.453 & VISTA/VIRCAM & $J$  & $21.159 \pm 0.229$ \\
0.499 & VISTA/VIRCAM & $K_s$ & $18.570 \pm 0.108$ \\
1.454 & VISTA/VIRCAM & $K_s$ & $17.834 \pm 0.048$ \\
2.450 & VISTA/VIRCAM & $K_s$ & $17.669 \pm 0.071$ \\
3.447 & VISTA/VIRCAM & $K_s$ & $17.575 \pm 0.038$ \\
4.446 & VISTA/VIRCAM & $K_s$ & $17.638 \pm 0.042$ \\
6.458 & VISTA/VIRCAM & $K_s$ & $17.859 \pm 0.041$ \\
7.449 & VISTA/VIRCAM & $K_s$ & $17.988 \pm 0.045$ \\
9.443 & VISTA/VIRCAM & $K_s$ & $18.528 \pm 0.060$ \\
10.445 & VISTA/VIRCAM & $K_s$ & $18.807 \pm 0.042$ \\
12.450 & VISTA/VIRCAM & $K_s$ & $19.416 \pm 0.139$ \\
14.452 & VISTA/VIRCAM & $K_s$ & $20.051 \pm 0.205$ \\
\enddata
\tablecomments{Column (1) gives the start time of the observation relative to the GW170817 gravitational-wave trigger. All observations were obtained with VISTA/VIRCAM with an exposure time of 120~s. Magnitudes are reported on the AB system and corrected for Milky Way foreground extinction adopting $E(B-V)=0.105$~mag \citep{schlafly2011}, corresponding to $A_Y=0.107$, $A_J=0.074$, and $A_{K_s}=0.032$~mag.}
\end{deluxetable}

\begin{figure*}[t]
\centering
    \includegraphics[width=0.8\textwidth]{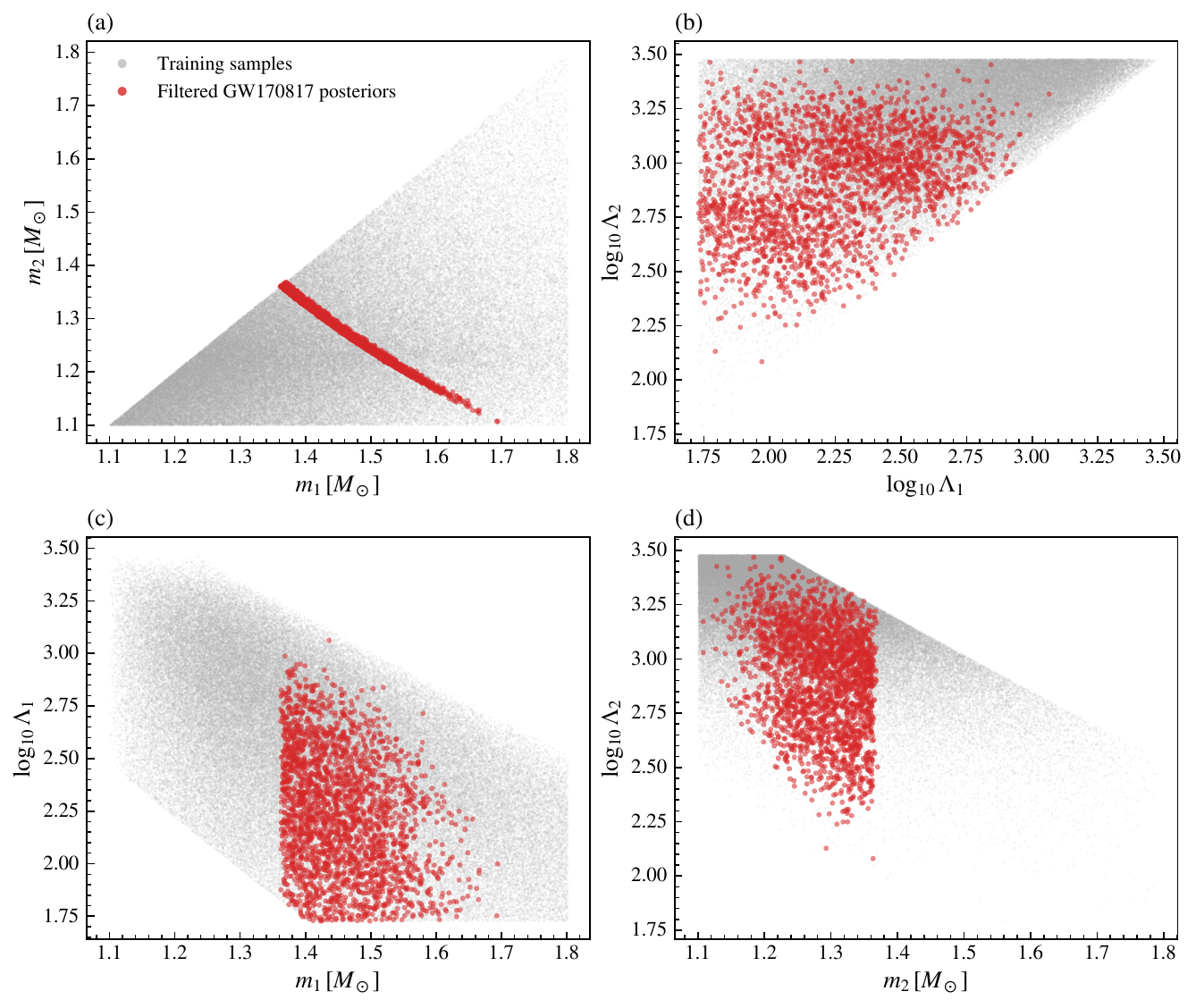}
    \caption{Distribution of the selected GW170817 posterior samples relative to the training parameter domain of the flow model. The gray points represent the training samples, and the red points show the GW170817 posterior samples remained in the constructed convex hull. The four panels show two dimension projections onto (a) $(m_1, m_2)$, (b) $(\log_{10}\Lambda_1,\log_{10}\Lambda_2)$, (c) $(m_1, \log_{10}\Lambda_1)$, and (d) $(m_2, \log_{10}\Lambda_2)$.}
    \label{fig:170817_posteriors}
\end{figure*}

\begin{figure*}[t]
\centering
    \includegraphics[width=\textwidth]{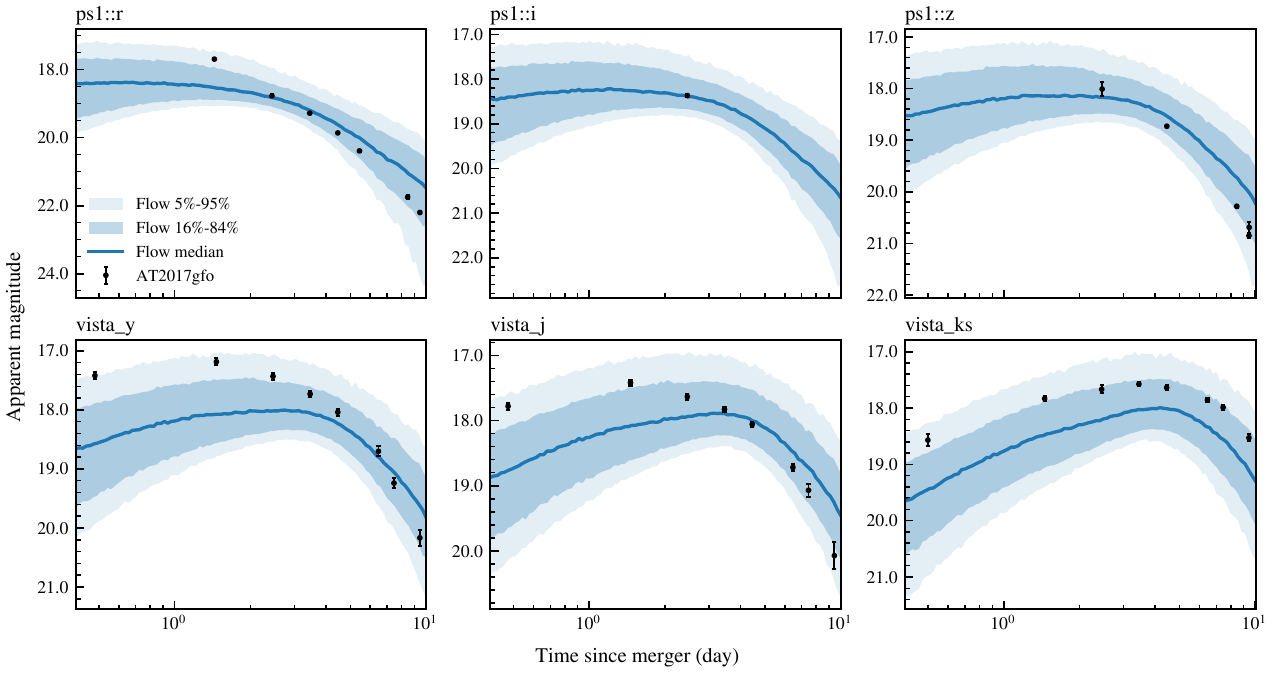}
    \caption{Flow prediction for GW170817/AT2017gfo. The solid blue lines show the median prediction from the flow, while the shaded region indicate the central 68\%, and 90\% quantile intervals. The points are the observations of AT2017gfo.}
    \label{fig:170817_flow}
\end{figure*}

As the GW posteriors span a broader parameter region than that covered by the flow training data, we restrict the posterior samples to remain within the training domain during prediction. For each posterior sample, the luminosity distance is first converted to redshift for transforming the component mass to the source frame. We then select only samples that lie within the four-dimensional convex hull constructed from the training conditions in $(m_1, m_2, \Lambda_1, \Lambda_2)$. The resulting filtered posteriors therefore form the approximation of the restricted distribution $\widetilde{q}(\boldsymbol{\vartheta})$ used in Eq.~\ref{eq:marginal2}. Construction of the convex hull is performed using the ConvexHull implementation in the \texttt{scipy.spatial} module from the \scipy package \citep{scipy20}. This procedure retains 2363 samples out of the original 8078 posterior samples. The selected posteriors are shown as red dots in Figure~\ref{fig:170817_posteriors}, along with the training distribution shown in gray. In particular, we take $\mu_{\mathrm{obs}} = |\cos\theta_{\rm JN}|$ to account for the viewing angle condition. These $(m_1, m_2, \Lambda_1, \Lambda_2, \mu_{\mathrm{obs}})$ samples are normalised using the same scaling procedure as in training and used as conditional inputs for the prediction. For each selected posterior sample, the flow model generates spectra at $100$ logarithmically spaced times between $0.4$ and $10$ days after merger. At each time, $4000$ latent samples are drawn to construct the predictive distribution. The luminosity distance associated with each posterior sample is used to convert the predicted spectra to the observer frame at a specific distance. The resulting multi-band light curves with the real observations are shown in Figure~\ref{fig:170817_flow}, where $\sim 77\%$ of the observations are well covered by the central 68\% predictive intervals of the flow model, indicating that the model can produce informative probabilistic light curve predictions using only GW posterior samples. 

\section{Discussion} \label{sec:discussion}

When \genova is applied to GW170817, the central 68\% predictive range covers $\sim 77\%$ of the observations in all bands, but does not fully enclose the earliest observations at $t\lesssim1.5$ day. By comparing this result with light curves generated directly from the underlying \nichollbns model using the same GW posterior samples, we find that this lack of early-time coverage is not unique to the surrogate prediction, but is already present in the underlying model distribution. This suggests that the early-time discrepancy is largely inherited from the adopted \nichollbns model, rather than being introduced primarily by the flow surrogate. We note that, although the training data include variations in the shocked-cocoon contribution and the parameter that enhances the early blue emission, these effects are marginalised over the adopted parameter ranges. The resulting central predictive range therefore captures the typical behaviour supported by the training distribution, which provides limited support for the most luminous early observations of AT2017gfo. 

This early-time behaviour should be viewed in the context of previous modelling of AT2017gfo. The early UV and optical observations revealed a luminous, rapidly evolving blue component within the first day after merger \citep[e.g.][]{dro17,Cow17,Eva17UV170817,Nicholl17}. This phase may be particularly sensitive to effects that are not fully captured by radiative-heating only prescription. These include possible ejecta geometric effects, additional luminosity sources such as wind-related blue emission and shock-cooling/cocoon emission \citep[e.g.][]{Per17,Pir18,Met20,nic21,hamidani24a}. In particular, within the model comparison of \citet{nic21}, the first day light curve of GW170817 strongly favours an additional luminosity source beyond the standard lanthanide-poor dynamical ejecta component, although the physical origin of this contribution is not unique \citep[e.g.,][]{Arc18}. 

The early and blue emission seen in AT2017gfo is also seen in the kilonova candidate associated with GRB 211211A, the only kilonova candidate with similarly early time observations. The blue excess in GRB 211211A is modelled by \citet{hamidani24b} to be consistent with late engine activity and subsequent interaction between the hot kilonova ejecta and a late jetted outflow.

These examples highlight a broader limitation: the current mapping from the binary parameters to the kilonova emission should only be understood within the assumptions of the adopted kilonova prescription. In the present implementation, this can be summarised as three main caveats.

\noindent\textit{1. Magnetar energy injection.}
This is related to the possible formation of a long-lived neutron star remnant. For systems that form a stable or supermassive remnant that rotates rapidly, the rotation energy can dominate over the original radioactive heating and produce a significantly different KN and the late-time afterglow \citep[e.g.,][]{sar22}. This effect is particularly relevant in the lower mass region (with the remnant mass $M_{\mathrm{rem}}\lesssim 1.2 M_{\rm TOV}$) in Figure~\ref{fig:170817_posteriors}. Although our current setup includes the blue enhancement effect from the magnetar wind and marginalises over $M_{\rm TOV}$, it does not explicitly include the additional energy injection from a magnetar remnant.

\noindent\textit{2. Restricted pre-merger spin configuration.}
A separate source of diversity arises from the pre-merger binary configuration. Although this work considers only low-spin scenarios, BNS systems with one highly spinning component can produce substantially more lanthanide-rich dynamical ejecta than corresponding irrotational systems, leading to brighter red kilonovae, which can constitute $\sim 4\%$ of the merging NS binaries \citep{ros24hs}. Though this effect may be partially degenerate with the dependence on mass ratio in existing numerical-relativity calibrated mappings, they are not expected to be completely covered. 

\noindent\textit{3. Kilonova-model and mapping systematics.}
A related limitation is that kilonova predictions are themselves model dependent. Current kilonova models differ not only in their computational approach (e.g., analytical, semi-analytical, or radiative transfer), but also in their physical assumptions such as ejecta geometry, composition, opacity treatment, radioactive heating rate, and thermalisation. Additional physical processes not included in many kilonova prescriptions, such as dust formation in the ejecta, could also affect the colour evolution of a KN through optical photon absorption and NIR re-emission \citep{tak14}. Although, dust is unlikely to dominate the NIR emission of AT2017gfo \citep{gall17}. These choices can lead to systematic variations in both the predicted observables and inferred results even for the same binary parameters, motivating $\gtrsim 1$ mag uncertainties in light curves \citep[e.g.,][]{Hei21,sar24b,Hus25}. Even within a fixed radiative-transfer framework, variations in atomic data and thermalisation efficiency prescriptions can produce substantial differences in both observables and inferred ejecta properties \citep[e.g.,][]{bu23,Bre24}. We therefore emphasise that the uncertainty represented in our probabilistic predictions reflects only the marginalised parameters within the adopted model, and does not fully capture the systematic uncertainty associated with the choice of kilonova prescription itself.

These results highlight the need for more comprehensive numerical simulations and a better understanding of kilonova model systematics. As kilonova models evolve, more physical effects that are not captured by the models considered here can be incorporated, including magnetar-driven energy injection, late-time engine activity, and spin effects. Improved numerical-relativity mappings between the BNS source parameters and ejecta properties may also reduce an important uncertainty in current kilonova predictions \citep[e.g.,][]{kru20,rad18,ned22,lof24}. In this context, \genova has the flexibility to be retained with alternative kilonova models or mixed training datasets spanning multiple model families,  while introducing separate modal labels as additional flow model conditions. In this way, this framework can not only consider uncertainty in the binary parameters, but also compare systematics from different models, and potentially incorporate predictions from multiple models in event-level analyses.

\section{Conclusions} \label{sec:conclusion}

We present \genova, a conditional normalising flow framework for probabilistic kilonova spectral prediction from GW-derived binary parameters. Conditioned only on rest-frame time, component masses, tidal deformabilities, and viewing angle, this framework can make probabilistic kilonova predictions with additional kilonova parameters marginalised. We validate its performance by comparing the median and central 68\% width ratio of the predictive distribution with the \nichollbns model. This presents the general reliability of the flow model between $0.4$--$10$ days after merger, with a relatively larger median error reaching $\sim 0.25$~mag for $u$~band after $1$~day, and deviations of the width ratio from unity of $\sim 20\%-80\%$ in the $K_s$ band before $1$~day. For other bands and epochs, the flow predictions remain consistent with the reference model, with the median residuals typically within $\sim 0.1$~mag and central width ratios within $\sim 20\%$ of unity. Since the flow predicts spectra rather than fixed multi-band light curves, it gives more flexibility for multi-band comparisons and observational applications in any filter of interest within the training wavelength range $2800$--$25000$~\ang in the rest frame. We also present newly reduced VISTA $Y$-, $J$-, and $K_s$-band photometry for comparison with the flow predictions. Applying \genova to GW170817 posterior samples demonstrates how the framework can propagate both GW parameter and nuisance parameter uncertainties into kilonova light curve predictions.

In this work, we adopt the \nichollbns model to train the normalising flow as a proof of concept for probabilistic kilonova prediction. The cross-model comparisons with \kasen illustrate that \genova can be used to identify regions of parameter space where different kilonova prescriptions produce consistent photometric behaviour. This motivates a future development of this framework in incorporating multiple kilonova prescriptions through additional conditioning labels. Such extensions support studying systematic difference between different KN models \citep[e.g.,][]{bu19, bu23}, while allowing model-specific predictive distributions to be compared or combined in one probabilistic analysis. Looking ahead, third-generation (3G) GW detectors are expected to detect BNS mergers at substantially higher rates and out to larger distances \citep[e.g.,][]{Mag20ET,Eva21CE}, electromagnetic follow-up will therefore benefit from frameworks that can propagate GW-inferred source properties efficiently into kilonova observables. In this context, \genova could serve as a downstream tool combined with rapid GW parameter estimation pipelines such as \textsc{Dingo}\citep{Dax21realtime,Dax25realtime} for probabilistic kilonova predictions from GW-detected BNS mergers. 

%%%%%%%%%%%%%%%%%%%%%%%%%%%%%%%%%%%
%%========acknowledgments=======%%%
%%%%%%%%%%%%%%%%%%%%%%%%%%%%%%%%%%%

\begin{acknowledgments}
We thank Thibeau Wouters and Peter P.H. Pang for helpful discussions on kilonova simulation. We also thank Lami Suleiman for useful discussions and suggestions on neutron star equation of state. XD is supported by Chinese Scholarship Council (CSC). GPL is supported by the Royal Society via a Dorothy Hodgkin Fellowship (Grant Nos. DHF-R1-221175 and DHF-ERE-221005). CM and ISH were supported by the Science and Technology Facilities Council (STFC) grants T/V005634/1 and UKRI2487. The Cosmic Dawn Center (DAWN) is funded by the Danish National Research Foundation under grant DNRF140. The observations with VISTA were gathered by the ESO VINROUGE Survey (198.D-2010). This work was supported by research grants (VIL16599,VIL54489) from VILLUM FONDEN. 
\end{acknowledgments}

\section*{Data Availability}
The full newly reduced VISTA photometry of AT2017gfo is available in this manuscript. The normalising flow framework with its source code, and all figure scripts of the paper will be released in a public repository after publication.

\facilities{ESO:VISTA(VIRCAM)}

\software{
\redback \citep{sar24a},
\gla \citep{mic24},
\nflows \citep{con20},
\pyt \citep{pas19}，
\sncosmo \citep{sncosmo},
\astropy \citep{astropy},
\sfft \citep{SFFT},
\photutils \citep{photutils},
\scipy \citep{scipy20}, 
\mat \citep{matplotlib},
\numpy \citep{numpy}.
}

%%%%%%%%%%%%%%%%%%%%%%%%%%%%%%%%%%%
%%===========appendix===========%%%
%%%%%%%%%%%%%%%%%%%%%%%%%%%%%%%%%%%
\appendix 
\restartappendixnumbering

\section{Flow configuration} \label{app:flow_config}

The spectra is trained for $3000$ epochs with a batch size of $500$. The architecture comprises $3$ coupling layers with $32$ neurons each. A learning rate scheduler (\texttt{CosAnnealingLR}) is applied for improved convergence, with an initial learning rate of $1\times 10^{-3}$ and a minimum learning rate of $1 \times 10^{-5}$. We adopted AdamW as the optimiser during the training. The architecture of the adopted flow model is summarised in Table~\ref{tab:flow_param}. Training was performed on a single NVIDIA GeForce RTX 5090 GPU and took approximately $10$ h wall-clock time.

\begin{deluxetable}{lc}[htbp]
\tablecaption{Configuration of the conditional normalising-flow model.\label{tab:flow_param}}
\tablewidth{\columnwidth}
\tabletypesize{\scriptsize}
\tablehead{
\colhead{Quantity} & \colhead{Adopted setting}
}
\startdata
Model type 
& Conditional RealNVP flow \\
Target variable 
& Rest-frame spectrum, $\mathbf{x}\in\mathbb{R}^{20}$ \\
Conditionals 
& $(t,m_1,m_2,\Lambda_1,\Lambda_2,\mu_{\rm obs})$ \\
Condition dimension 
& 6 \\
Latent distribution 
& $\mathcal{N}(\mathbf{0},\mathbf{I})$ \\
Coupling transforms 
& 3 \\
Hidden neurons 
& 32 \\
Activation function 
& ReLU \\
Training objective 
& Negative log-likelihood \\
Optimiser 
& AdamW \\
Initial learning rate 
& $1\times10^{-3}$ \\
Scheduler 
& \texttt{CosineAnnealingLR} \\
Minimum learning rate 
& $1\times10^{-5}$ \\
Batch size 
& 500 \\
Training epoch
& 3000 \\
\enddata
\tablecomments{
The remaining kilonova parameters are varied during data generation but are not included in the conditional parameters, their effect is absorbed implicitly into the learned conditional data distribution.}
\end{deluxetable} 

\clearpage
\section{All bands plots} \label{app:test_plots}
\begin{figure*}[ht!]
\centering
    \includegraphics[width=1.0\textwidth]{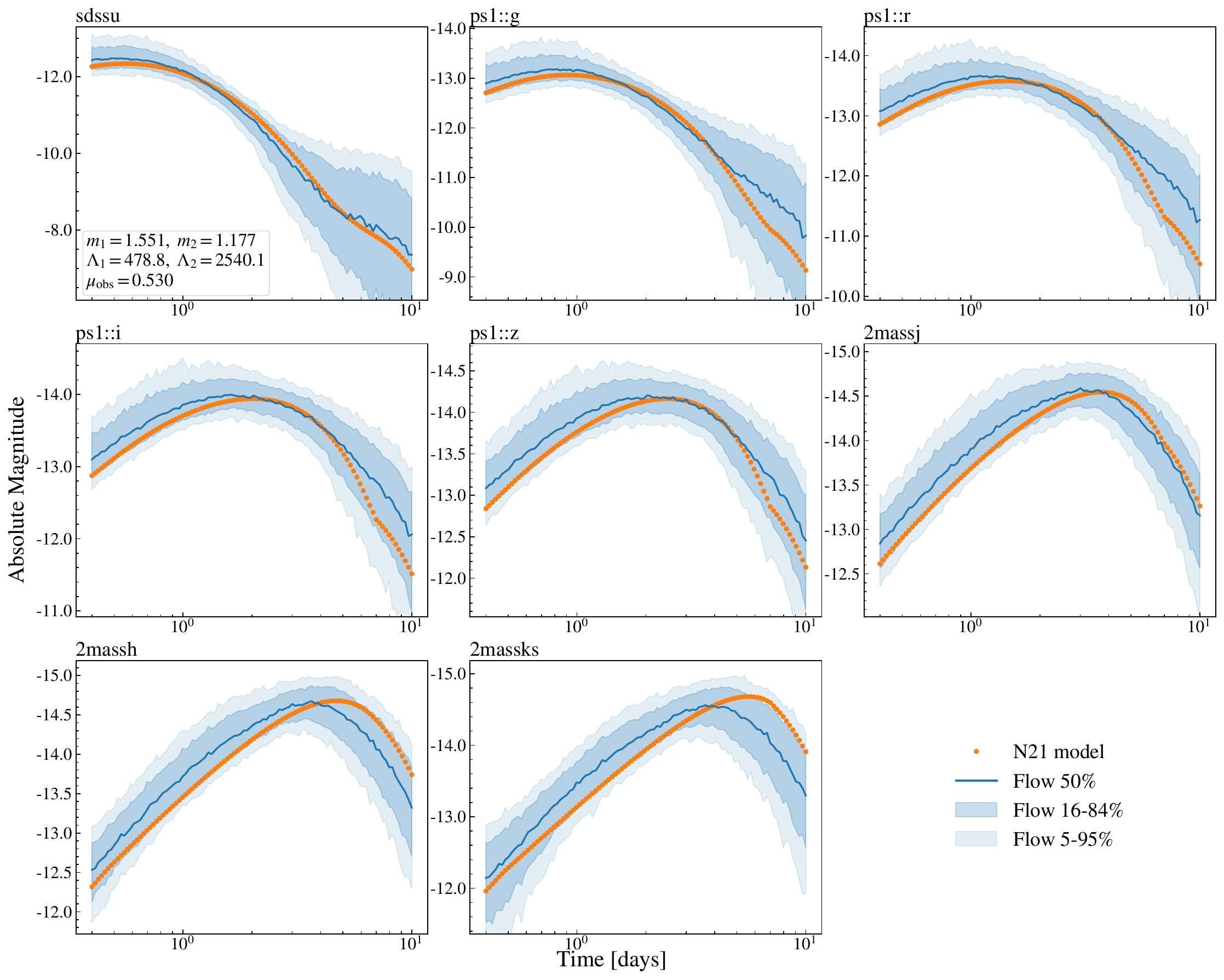}
    \caption{Multi-band light curve comparison for the in-distribution testing example. The predicted multi-band light curves from the flow model are shown by the median (blue solid line) and central 68\% quantile interval (blue shaded region), compared with the theoretical \nichollbns light curve (orange points). The $(m,\Lambda)$ values for this example are drawn from the testing distribution.}
    \label{fig:test_bands}
\end{figure*}

\begin{figure*}[ht!]
\centering
    \includegraphics[width=1.0\textwidth]{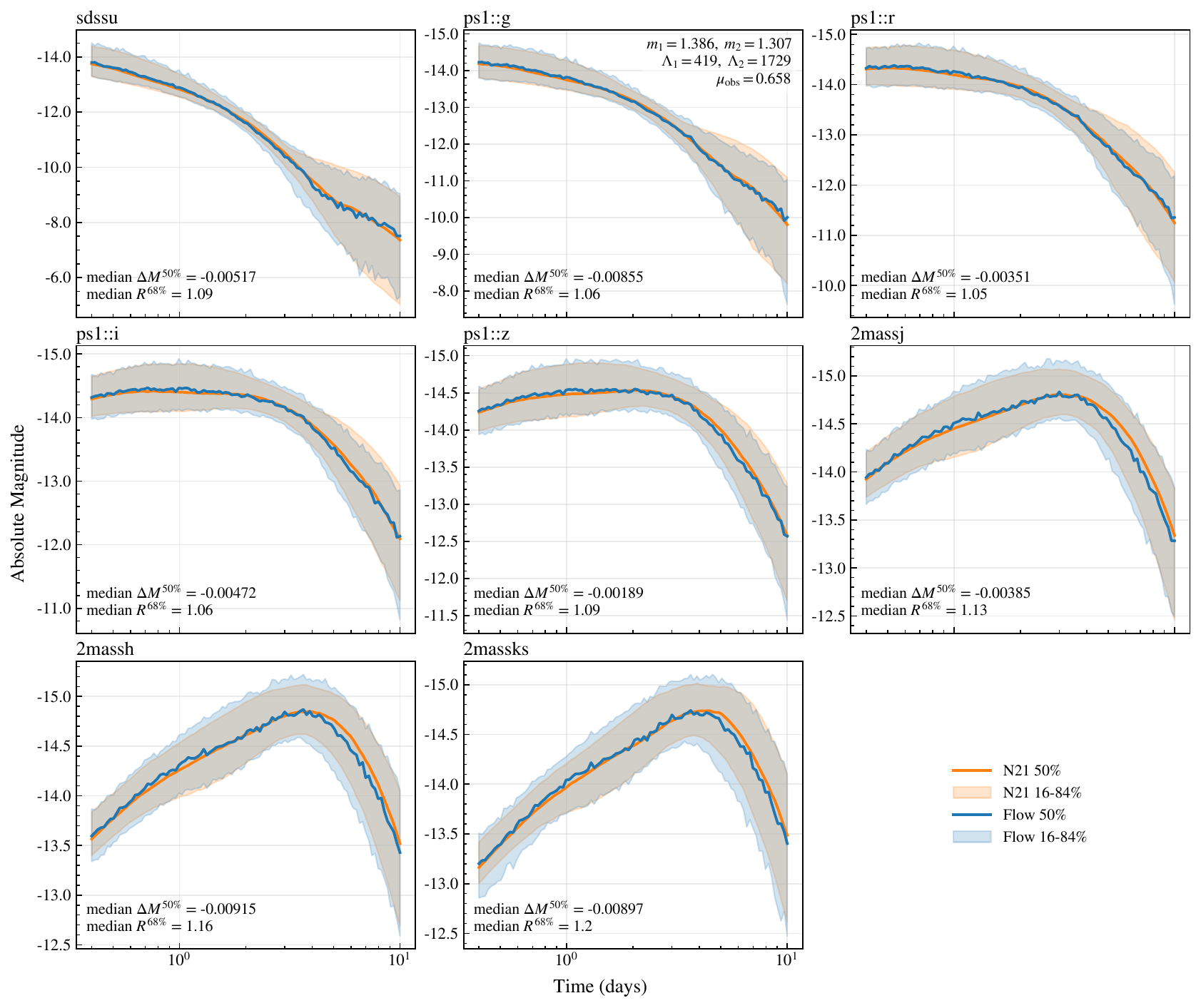}
    \caption{Multi-band light curve comparison between the \nichollbns and the flow model for an example test case. The median and central 68\% quantile interval are shown for both models (flow: blue; theory: orange). The average median residual and the ratio of uncertainty width over time for each band are shown at the left bottom corner in each panel.}
    \label{fig:theory_bands}
\end{figure*}

Figure~\ref{fig:test_bands} show the multi-band light curve predictions from the flow model for a test example. The predictions are shown against one corresponding example test light curve. Figure~\ref{fig:theory_bands} presents the light curve distribution comparison between the flow prediction and \nichollbns using one example set of test parameters. 

\clearpage
\section{VISTA image subtraction and photometry}
\label{app:vista_reduction}

The $Y$-, $J$-, and $K_s$-band VISTA/VIRCAM observations of AT2017gfo were re-analysed using late-time imaging of NGC~4993, obtained after the transient had faded, as reference images. The late-time observations were obtained on 2019 March 19, with three tiled images acquired in each of the $Y$, $J$, and $K_s$ bands, each with an exposure time of 240~s. The observations spanned 02:38:35--04:48:04~UTC. Starting from the reduced and astrometrically aligned science and reference images, the subtraction and photometric measurements were performed independently in each band. The final resampled images have an effective pixel scale of approximately $0.341$~arcsec~pixel$^{-1}$, as determined from their astrometric solutions.

Prior to image subtraction, low-spatial-frequency background variations in both the science and reference images were modelled and removed using the \texttt{Background2D} implementation in \photutils \citep{photutils}. The background was estimated using 48-pixel boxes, with a $5\times5$ median filter applied to the background mesh and iterative $2.5\sigma$ clipping. The transient position was masked during the background estimation to prevent source flux from influencing the model.

For each science epoch, five isolated field stars were copied to otherwise blank regions of the science image before subtraction. Each copied source was inserted using a $57\times57$ pixel stamp, after removing the median local background measured around the original star and matching the stamp pedestal to the background at the destination position. This provided a set of calibration sources that underwent the same convolution and subtraction procedure as the transient. The transient and host-galaxy region, together with the copied stars, were excluded from the source sample used to determine the subtraction solution.

Difference images were produced using the \sfft image-subtraction algorithm \citep{SFFT}. For each epoch, the reference image was convolved to match the corresponding science image. The convolution kernel was allowed to vary spatially as a second-order polynomial, while a spatially constant photometric scaling was adopted. The kernel half-width was determined automatically, with the allowed range restricted to 2--20 pixels.

Following image subtraction, the remaining smooth residual associated with NGC~4993 was modelled locally within a radius of 140 pixels around the galaxy centre using a 15-pixel median filter. The transient position was masked while constructing this model, which was then subtracted to produce the final image used for photometry.

Circular-aperture photometry was performed on the final difference images using an aperture radius of 3 pixels. The transient position was determined using \texttt{DAOStarFinder} from \photutils \citep{photutils} within a restricted region around its expected location. A detected centroid was accepted only when it lay within 4 pixels of the expected position; otherwise, including at low signal-to-noise epochs, forced photometry was performed at the expected transient coordinates.

The local background was estimated using circular sky apertures with the same radius as the source aperture. These were positioned separately for each epoch and were adjusted interactively to avoid compact subtraction artefacts and structured residual emission from NGC~4993. For a source aperture containing a summed signal $S_{\rm tr}$, the background-subtracted transient counts were calculated as
\begin{equation}
    F_{\rm tr}
    =
    S_{\rm tr}
    -
    A_{\rm ap}
    \left\langle \overline{B}_j \right\rangle,
    \label{eq:transient_net_counts}
\end{equation}
where $A_{\rm ap}$ is the area of the source aperture, $\overline{B}_j$ is the mean pixel value in the $j$th sky aperture, and the angle brackets denote the mean over the selected sky apertures.

The same aperture radius was used to measure the five copied calibration stars. Their centroids were determined independently at each epoch and their local backgrounds were estimated using surrounding sky apertures. Each calibration star therefore provided an independent estimate of the transient magnitude,
\begin{equation}
    m_{b,i}=m_{b,\star i}-2.5\log_{10}\left(\frac{F_{\rm tr}}{F_{\star i}}\right)+\Delta_b,
\end{equation}
where $m_{b,\star i}$ and $F_{\star i}$ are the catalogue magnitude and measured background-subtracted counts of calibration star $i$ in band $b$. The catalogue magnitudes are on the VISTA Vega system and were converted to the AB system using the VISTA/VIRCAM offsets of \citet{gonzalezfernandez2018}, $Y_{\rm AB}-Y_{\rm Vega}=0.600$, $J_{\rm AB}-J_{\rm Vega}=0.916$, and $K_{s,\rm AB}-K_{s,\rm Vega}=1.827$~mag.

The photometry was subsequently corrected for Milky Way foreground extinction. We adopt $E(B-V)=0.105$~mag for the line of sight to AT2017gfo, following the recalibrated Galactic dust maps of \citet{schlafly2011}. Assuming $R_V=3.1$, the VISTA/VIRCAM extinction coefficients $A_Y/E(B-V)=1.017$, $A_J/E(B-V)=0.705$, and $A_{K_s}/E(B-V)=0.308$ from \citet{gonzalezfernandez2018} give $A_Y=0.107$, $A_J=0.074$, and $A_{K_s}=0.032$~mag. The reported magnitude at each epoch was taken as the unweighted mean of the valid measurements obtained from the five calibration stars.

The empirical transient-flux uncertainty was estimated from the dispersion of the integrated counts measured in the target sky apertures,
\begin{equation}
    \sigma_{m,{\rm tr}}
    =
    \frac{2.5}{\ln 10}
    \frac{\sigma_{F,{\rm sky}}}{F_{\rm tr}},
    \qquad
    \sigma_{F,{\rm sky}}
    =
    \left[
    \mathrm{Var}
    \left(
    S_{{\rm sky},j}
    \right)
    \right]^{1/2},
    \label{eq:transient_magnitude_uncertainty}
\end{equation}
where $S_{{\rm sky},j}$ is the summed signal in the $j$th sky aperture. For each calibration star, this term was combined in quadrature with the corresponding catalogue-magnitude uncertainty. The final uncertainty on the reported magnitude was calculated as
\begin{equation}
    \sigma_{m_b}
    =
    \left[
    s^2(m_{b,i})
    +
    \left\langle
    \sigma_{m_{b,i}}
    \right\rangle^2
    \right]^{1/2},
    \label{eq:mean_photometric_uncertainty}
\end{equation}
where $s(m_{b,i})$ is the sample standard deviation of the independent star-calibrated magnitudes and $\langle\sigma_{m_{b,i}}\rangle$ is their mean formal uncertainty.

\clearpage
\bibliography{kilonova}{}

@ARTICLE{hamidani24a,
       author = {{Hamidani}, Hamid and {Kimura}, Shigeo S. and {Tanaka}, Masaomi and {Ioka}, Kunihito},
        title = "{Late Engine Activity in Neutron Star Mergers and Its Cocoon: An Alternative Scenario for the Blue Kilonova}",
      journal = {\apj},
         year = 2024,
        month = mar,
       volume = {963},
       number = {2},
          eid = {137},
        pages = {137},
          doi = {10.3847/1538-4357/ad20d0},
archivePrefix = {arXiv},
       eprint = {2312.06286},
 primaryClass = {astro-ph.HE},
       adsurl = {https://ui.adsabs.harvard.edu/abs/2024ApJ...963..137H}
}

@ARTICLE{hamidani24b,
       author = {{Hamidani}, Hamid and {Tanaka}, Masaomi and {Kimura}, Shigeo S. and {Lamb}, Gavin P. and {Kawaguchi}, Kyohei},
        title = "{GRB 211211A: The Case for an Engine-powered over r-process-powered Blue Kilonova}",
      journal = {\apjl},
         year = 2024,
        month = aug,
       volume = {971},
       number = {2},
          eid = {L30},
        pages = {L30},
          doi = {10.3847/2041-8213/ad6864},
archivePrefix = {arXiv},
       eprint = {2406.14366},
 primaryClass = {astro-ph.HE},
       adsurl = {https://ui.adsabs.harvard.edu/abs/2024ApJ...971L..30H}
}

@article{nic21,
    author = {Nicholl, Matt and Margalit, Ben and Schmidt, Patricia and Smith, Graham P and Ridley, Evan J and Nuttall, James},
    title = {Tight multimessenger constraints on the neutron star equation of state from GW170817 and a forward model for kilonova light-curve synthesis},
    journal = {Monthly Notices of the Royal Astronomical Society},
    volume = {505},
    number = {2},
    pages = {3016-3032},
    year = {2021},
    month = {05},
    issn = {0035-8711},
    doi = {10.1093/mnras/stab1523},
    url = {https://doi.org/10.1093/mnras/stab1523},
    eprint = {https://academic.oup.com/mnras/article-pdf/505/2/3016/39571575/stab1523.pdf},
}

@ARTICLE{sar22,
       author = {{Sarin}, Nikhil and {Omand}, Conor M.~B. and {Margalit}, Ben and {Jones}, David I.},
        title = "{On the diversity of magnetar-driven kilonovae}",
      journal = {\mnras},
         year = 2022,
        month = nov,
       volume = {516},
       number = {4},
        pages = {4949-4962},
          doi = {10.1093/mnras/stac2609},
archivePrefix = {arXiv},
       eprint = {2205.14159},
 primaryClass = {astro-ph.HE},
       adsurl = {https://ui.adsabs.harvard.edu/abs/2022MNRAS.516.4949S}
}

@article{sar24a,
    author = {Sarin, Nikhil and H{\"u}bner, Moritz and Omand, Conor M B and Setzer, Christian N and Schulze, Steve and Adhikari, Naresh and Sagu{\'e}s-Carracedo, Ana and Galaudage, Shanika and Wallace, Wendy F and Lamb, Gavin P and Lin, En-Tzu},
    title = {redback: a Bayesian inference software package for electromagnetic transients},
    journal = {Monthly Notices of the Royal Astronomical Society},
    volume = {531},
    number = {1},
    pages = {1203-1227},
    year = {2024},
    month = {05},
    issn = {0035-8711},
    doi = {10.1093/mnras/stae1238},
    url = {https://doi.org/10.1093/mnras/stae1238},
    eprint = {https://academic.oup.com/mnras/article-pdf/531/1/1203/57801464/stae1238.pdf},
}

@article{sar24b,
doi = {10.3847/2041-8213/ad739d},
url = {https://doi.org/10.3847/2041-8213/ad739d},
year = {2024},
month = {sep},
publisher = {The American Astronomical Society},
volume = {973},
number = {1},
pages = {L24},
author = {Sarin, Nikhil and Rosswog, Stephan},
title = {Cautionary Tales on Heating-rate Prescriptions in Kilonovae},
journal = {The Astrophysical Journal Letters}
}

@ARTICLE{ros24hs,
       author = {{Rosswog}, S. and {Diener}, P. and {Torsello}, F. and {Tauris}, T.~M. and {Sarin}, N.},
        title = "{Mergers of double NSs with one high-spin component: brighter kilonovae and fallback accretion, weaker gravitational waves}",
      journal = {\mnras},
         year = 2024,
        month = may,
       volume = {530},
       number = {2},
        pages = {2336-2354},
          doi = {10.1093/mnras/stae454},
archivePrefix = {arXiv},
       eprint = {2310.15920},
 primaryClass = {astro-ph.HE},
       adsurl = {https://ui.adsabs.harvard.edu/abs/2024MNRAS.530.2336R}
}

@article{Die17,
doi = {10.1088/1361-6382/aa6bb0},
url = {https://dx.doi.org/10.1088/1361-6382/aa6bb0},
year = {2017},
month = {apr},
publisher = {IOP Publishing},
volume = {34},
number = {10},
pages = {105014},
author = {Tim Dietrich and Maximiliano Ujevic},
title = {Modeling dynamical ejecta from binary neutron star mergers and implications for electromagnetic counterparts},
journal = {Classical and Quantum Gravity}
}

@article{cou19,
    author = {Coughlin, Michael W and Dietrich, Tim and Margalit, Ben and Metzger, Brian D},
    title = "{Multimessenger Bayesian parameter inference of a binary neutron star merger}",
    journal = {Monthly Notices of the Royal Astronomical Society: Letters},
    volume = {489},
    number = {1},
    pages = {L91-L96},
    year = {2019},
    month = {08},
    issn = {1745-3925},
    doi = {10.1093/mnrasl/slz133},
    url = {https://doi.org/10.1093/mnrasl/slz133},
    eprint = {https://academic.oup.com/mnrasl/article-pdf/489/1/L91/56978543/mnrasl\_489\_1\_l91.pdf},
}

@article{rad18,
doi = {10.3847/1538-4357/aaf054},
url = {https://dx.doi.org/10.3847/1538-4357/aaf054},
year = {2018},
month = {dec},
publisher = {The American Astronomical Society},
volume = {869},
number = {2},
pages = {130},
author = {David Radice and Albino Perego and Kenta Hotokezaka and Steven A. Fromm and Sebastiano Bernuzzi and Luke F. Roberts},
title = {Binary Neutron Star Mergers: Mass Ejection, Electromagnetic Counterparts, and Nucleosynthesis},
journal = {The Astrophysical Journal}
}

@ARTICLE{rad18EOS,
       author = {{Radice}, David and {Perego}, Albino and {Zappa}, Francesco and {Bernuzzi}, Sebastiano},
        title = "{GW170817: Joint Constraint on the Neutron Star Equation of State from Multimessenger Observations}",
      journal = {\apjl},
         year = 2018,
        month = jan,
       volume = {852},
       number = {2},
          eid = {L29},
        pages = {L29},
          doi = {10.3847/2041-8213/aaa402},
archivePrefix = {arXiv},
       eprint = {1711.03647},
 primaryClass = {astro-ph.HE},
       adsurl = {https://ui.adsabs.harvard.edu/abs/2018ApJ...852L..29R}
}

@ARTICLE{kas17,
       author = {Kasen, Daniel and {Metzger}, Brian and {Barnes}, Jennifer and {Quataert}, Eliot and {Ramirez-Ruiz}, Enrico},
        title = "{Origin of the heavy elements in binary neutron-star mergers from a gravitational-wave event}",
      journal = {\nat},
         year = 2017,
        month = nov,
       volume = {551},
       number = {7678},
        pages = {80-84},
          doi = {10.1038/nature24453},
archivePrefix = {arXiv},
       eprint = {1710.05463},
 primaryClass = {astro-ph.HE},
       adsurl = {https://ui.adsabs.harvard.edu/abs/2017Natur.551...80K}
}

@article{ros24hr,
author = {Rosswog, Stephan and Korobkin, Oleg},
title = {Heavy Elements and Electromagnetic Transients from Neutron Star Mergers},
journal = {Annalen der Physik},
volume = {536},
number = {2},
pages = {2200306},
doi = {https://doi.org/10.1002/andp.202200306},
url = {https://onlinelibrary.wiley.com/doi/abs/10.1002/andp.202200306},
eprint = {https://onlinelibrary.wiley.com/doi/pdf/10.1002/andp.202200306},
year = {2024}
}

@ARTICLE{vil17,
       author = {{Villar}, V.~A. and {Guillochon}, J. and {Berger}, E. and {Metzger}, B.~D. and {Cowperthwaite}, P.~S. and {Nicholl}, M. and {Alexander}, K.~D. and {Blanchard}, P.~K. and {Chornock}, R. and {Eftekhari}, T. and {Fong}, W. and {Margutti}, R. and {Williams}, P.~K.~G.},
        title = "{The Combined Ultraviolet, Optical, and Near-infrared Light Curves of the Kilonova Associated with the Binary Neutron Star Merger GW170817: Unified Data Set, Analytic Models, and Physical Implications}",
      journal = {\apjl},
         year = 2017,
        month = dec,
       volume = {851},
       number = {1},
          eid = {L21},
        pages = {L21},
          doi = {10.3847/2041-8213/aa9c84},
archivePrefix = {arXiv},
       eprint = {1710.11576},
 primaryClass = {astro-ph.HE},
       adsurl = {https://ui.adsabs.harvard.edu/abs/2017ApJ...851L..21V}
}

@article{abb17a,
  title = {GW170817: Observation of Gravitational Waves from a Binary Neutron Star Inspiral},
  author = {Abbott, B. P. and Abbott, R. and Abbott, T. D. and Acernese, F. and Ackley, K. and Adams, C. and Adams, T. and Addesso, P. and Adhikari, R. X. and Adya, V. B. and Affeldt, C. and Afrough, M. and Agarwal, B. and Agathos, M. and Agatsuma, K. and Aggarwal, N. and Aguiar, O. D. and Aiello, L. and Ain, A. and Ajith, P. and Allen, B. and Allen, G. and Allocca, A. and Altin, P. A. and Amato, A. and Ananyeva, A. and Anderson, S. B. and Anderson, W. G. and Angelova, S. V. and Antier, S. and Appert, S. and Arai, K. and Araya, M. C. and Areeda, J. S. and Arnaud, N. and Arun, K. G. and Ascenzi, S. and Ashton, G. and Ast, M. and Aston, S. M. and Astone, P. and Atallah, D. V. and Aufmuth, P. and Aulbert, C. and AultONeal, K. and Austin, C. and Avila-Alvarez, A. and Babak, S. and Bacon, P. and Bader, M. K. M. and Bae, S. and Bailes, M. and Baker, P. T. and Baldaccini, F. and Ballardin, G. and Ballmer, S. W. and Banagiri, S. and Barayoga, J. C. and Barclay, S. E. and Barish, B. C. and Barker, D. and Barkett, K. and Barone, F. and Barr, B. and Barsotti, L. and Barsuglia, M. and Barta, D. and Barthelmy, S. D. and Bartlett, J. and Bartos, I. and Bassiri, R. and Basti, A. and Batch, J. C. and Bawaj, M. and Bayley, J. C. and Bazzan, M. and B\'ecsy, B. and Beer, C. and Bejger, M. and Belahcene, I. and Bell, A. S. and Berger, B. K. and Bergmann, G. and Bernuzzi, S. and Bero, J. J. and Berry, C. P. L. and Bersanetti, D. and Bertolini, A. and Betzwieser, J. and Bhagwat, S. and Bhandare, R. and Bilenko, I. A. and Billingsley, G. and Billman, C. R. and Birch, J. and Birney, I. A. and Birnholtz, O. and Biscans, S. and Biscoveanu, S. and Bisht, A. and Bitossi, M. and Biwer, C. and Bizouard, M. A. and Blackburn, J. K. and Blackman, J. and Blair, C. D. and Blair, D. G. and Blair, R. M. and Bloemen, S. and Bock, O. and Bode, N. and Boer, M. and Bogaert, G. and Bohe, A. and Bondu, F. and Bonilla, E. and Bonnand, R. and Boom, B. A. and Bork, R. and Boschi, V. and Bose, S. and Bossie, K. and Bouffanais, Y. and Bozzi, A. and Bradaschia, C. and Brady, P. R. and Branchesi, M. and Brau, J. E. and Briant, T. and Brillet, A. and Brinkmann, M. and Brisson, V. and Brockill, P. and Broida, J. E. and Brooks, A. F. and Brown, D. A. and Brown, D. D. and Brunett, S. and Buchanan, C. C. and Buikema, A. and Bulik, T. and Bulten, H. J. and Buonanno, A. and Buskulic, D. and Buy, C. and Byer, R. L. and Cabero, M. and Cadonati, L. and Cagnoli, G. and Cahillane, C. and Calder\'on Bustillo, J. and Callister, T. A. and Calloni, E. and Camp, J. B. and Canepa, M. and Canizares, P. and Cannon, K. C. and Cao, H. and Cao, J. and Capano, C. D. and Capocasa, E. and Carbognani, F. and Caride, S. and Carney, M. F. and Carullo, G. and Casanueva Diaz, J. and Casentini, C. and Caudill, S. and Cavagli\`a, M. and Cavalier, F. and Cavalieri, R. and Cella, G. and Cepeda, C. B. and Cerd\'a-Dur\'an, P. and Cerretani, G. and Cesarini, E. and Chamberlin, S. J. and Chan, M. and Chao, S. and Charlton, P. and Chase, E. and Chassande-Mottin, E. and Chatterjee, D. and Chatziioannou, K. and Cheeseboro, B. D. and Chen, H. Y. and Chen, X. and Chen, Y. and Cheng, H.-P. and Chia, H. and Chincarini, A. and Chiummo, A. and Chmiel, T. and Cho, H. S. and Cho, M. and Chow, J. H. and Christensen, N. and Chu, Q. and Chua, A. J. K. and Chua, S. and Chung, A. K. W. and Chung, S. and Ciani, G. and Ciolfi, R. and Cirelli, C. E. and Cirone, A. and Clara, F. and Clark, J. A. and Clearwater, P. and Cleva, F. and Cocchieri, C. and Coccia, E. and Cohadon, P.-F. and Cohen, D. and Colla, A. and Collette, C. G. and Cominsky, L. R. and Constancio, M. and Conti, L. and Cooper, S. J. and Corban, P. and Corbitt, T. R. and Cordero-Carri\'on, I. and Corley, K. R. and Cornish, N. and Corsi, A. and Cortese, S. and Costa, C. A. and Coughlin, M. W. and Coughlin, S. B. and Coulon, J.-P. and Countryman, S. T. and Couvares, P. and Covas, P. B. and Cowan, E. E. and Coward, D. M. and Cowart, M. J. and Coyne, D. C. and Coyne, R. and Creighton, J. D. E. and Creighton, T. D. and Cripe, J. and Crowder, S. G. and Cullen, T. J. and Cumming, A. and Cunningham, L. and Cuoco, E. and Dal Canton, T. and D\'alya, G. and Danilishin, S. L. and D'Antonio, S. and Danzmann, K. and Dasgupta, A. and Da Silva Costa, C. F. and Dattilo, V. and Dave, I. and Davier, M. and Davis, D. and Daw, E. J. and Day, B. and De, S. and DeBra, D. and Degallaix, J. and De Laurentis, M. and Del\'eglise, S. and Del Pozzo, W. and Demos, N. and Denker, T. and Dent, T. and De Pietri, R. and Dergachev, V. and De Rosa, R. and DeRosa, R. T. and De Rossi, C. and DeSalvo, R. and de Varona, O. and Devenson, J. and Dhurandhar, S. and D\'{\i}az, M. C. and Dietrich, T. and Di Fiore, L. and Di Giovanni, M. and Di Girolamo, T. and Di Lieto, A. and Di Pace, S. and Di Palma, I. and Di Renzo, F. and Doctor, Z. and Dolique, V. and Donovan, F. and Dooley, K. L. and Doravari, S. and Dorrington, I. and Douglas, R. and Dovale \'Alvarez, M. and Downes, T. P. and Drago, M. and Dreissigacker, C. and Driggers, J. C. and Du, Z. and Ducrot, M. and Dudi, R. and Dupej, P. and Dwyer, S. E. and Edo, T. B. and Edwards, M. C. and Effler, A. and Eggenstein, H.-B. and Ehrens, P. and Eichholz, J. and Eikenberry, S. S. and Eisenstein, R. A. and Essick, R. C. and Estevez, D. and Etienne, Z. B. and Etzel, T. and Evans, M. and Evans, T. M. and Factourovich, M. and Fafone, V. and Fair, H. and Fairhurst, S. and Fan, X. and Farinon, S. and Farr, B. and Farr, W. M. and Fauchon-Jones, E. J. and Favata, M. and Fays, M. and Fee, C. and Fehrmann, H. and Feicht, J. and Fejer, M. M. and Fernandez-Galiana, A. and Ferrante, I. and Ferreira, E. C. and Ferrini, F. and Fidecaro, F. and Finstad, D. and Fiori, I. and Fiorucci, D. and Fishbach, M. and Fisher, R. P. and Fitz-Axen, M. and Flaminio, R. and Fletcher, M. and Fong, H. and Font, J. A. and Forsyth, P. W. F. and Forsyth, S. S. and Fournier, J.-D. and Frasca, S. and Frasconi, F. and Frei, Z. and Freise, A. and Frey, R. and Frey, V. and Fries, E. M. and Fritschel, P. and Frolov, V. V. and Fulda, P. and Fyffe, M. and Gabbard, H. and Gadre, B. U. and Gaebel, S. M. and Gair, J. R. and Gammaitoni, L. and Ganija, M. R. and Gaonkar, S. G. and Garcia-Quiros, C. and Garufi, F. and Gateley, B. and Gaudio, S. and Gaur, G. and Gayathri, V. and Gehrels, N. and Gemme, G. and Genin, E. and Gennai, A. and George, D. and George, J. and Gergely, L. and Germain, V. and Ghonge, S. and Ghosh, Abhirup and Ghosh, Archisman and Ghosh, S. and Giaime, J. A. and Giardina, K. D. and Giazotto, A. and Gill, K. and Glover, L. and Goetz, E. and Goetz, R. and Gomes, S. and Goncharov, B. and Gonz\'alez, G. and Gonzalez Castro, J. M. and Gopakumar, A. and Gorodetsky, M. L. and Gossan, S. E. and Gosselin, M. and Gouaty, R. and Grado, A. and Graef, C. and Granata, M. and Grant, A. and Gras, S. and Gray, C. and Greco, G. and Green, A. C. and Gretarsson, E. M. and Groot, P. and Grote, H. and Grunewald, S. and Gruning, P. and Guidi, G. M. and Guo, X. and Gupta, A. and Gupta, M. K. and Gushwa, K. E. and Gustafson, E. K. and Gustafson, R. and Halim, O. and Hall, B. R. and Hall, E. D. and Hamilton, E. Z. and Hammond, G. and Haney, M. and Hanke, M. M. and Hanks, J. and Hanna, C. and Hannam, M. D. and Hannuksela, O. A. and Hanson, J. and Hardwick, T. and Harms, J. and Harry, G. M. and Harry, I. W. and Hart, M. J. and Haster, C.-J. and Haughian, K. and Healy, J. and Heidmann, A. and Heintze, M. C. and Heitmann, H. and Hello, P. and Hemming, G. and Hendry, M. and Heng, I. S. and Hennig, J. and Heptonstall, A. W. and Heurs, M. and Hild, S. and Hinderer, T. and Ho, W. C. G. and Hoak, D. and Hofman, D. and Holt, K. and Holz, D. E. and Hopkins, P. and Horst, C. and Hough, J. and Houston, E. A. and Howell, E. J. and Hreibi, A. and Hu, Y. M. and Huerta, E. A. and Huet, D. and Hughey, B. and Husa, S. and Huttner, S. H. and Huynh-Dinh, T. and Indik, N. and Inta, R. and Intini, G. and Isa, H. N. and Isac, J.-M. and Isi, M. and Iyer, B. R. and Izumi, K. and Jacqmin, T. and Jani, K. and Jaranowski, P. and Jawahar, S. and Jim\'enez-Forteza, F. and Johnson, W. W. and Johnson-McDaniel, N. K. and Jones, D. I. and Jones, R. and Jonker, R. J. G. and Ju, L. and Junker, J. and Kalaghatgi, C. V. and Kalogera, V. and Kamai, B. and Kandhasamy, S. and Kang, G. and Kanner, J. B. and Kapadia, S. J. and Karki, S. and Karvinen, K. S. and Kasprzack, M. and Kastaun, W. and Katolik, M. and Katsavounidis, E. and Katzman, W. and Kaufer, S. and Kawabe, K. and K\'ef\'elian, F. and Keitel, D. and Kemball, A. J. and Kennedy, R. and Kent, C. and Key, J. S. and Khalili, F. Y. and Khan, I. and Khan, S. and Khan, Z. and Khazanov, E. A. and Kijbunchoo, N. and Kim, Chunglee and Kim, J. C. and Kim, K. and Kim, W. and Kim, W. S. and Kim, Y.-M. and Kimbrell, S. J. and King, E. J. and King, P. J. and Kinley-Hanlon, M. and Kirchhoff, R. and Kissel, J. S. and Kleybolte, L. and Klimenko, S. and Knowles, T. D. and Koch, P. and Koehlenbeck, S. M. and Koley, S. and Kondrashov, V. and Kontos, A. and Korobko, M. and Korth, W. Z. and Kowalska, I. and Kozak, D. B. and Kr\"amer, C. and Kringel, V. and Krishnan, B. and Kr\'olak, A. and Kuehn, G. and Kumar, P. and Kumar, R. and Kumar, S. and Kuo, L. and Kutynia, A. and Kwang, S. and Lackey, B. D. and Lai, K. H. and Landry, M. and Lang, R. N. and Lange, J. and Lantz, B. and Lanza, R. K. and Larson, S. L. and Lartaux-Vollard, A. and Lasky, P. D. and Laxen, M. and Lazzarini, A. and Lazzaro, C. and Leaci, P. and Leavey, S. and Lee, C. H. and Lee, H. K. and Lee, H. M. and Lee, H. W. and Lee, K. and Lehmann, J. and Lenon, A. and Leon, E. and Leonardi, M. and Leroy, N. and Letendre, N. and Levin, Y. and Li, T. G. F. and Linker, S. D. and Littenberg, T. B. and Liu, J. and Liu, X. and Lo, R. K. L. and Lockerbie, N. A. and London, L. T. and Lord, J. E. and Lorenzini, M. and Loriette, V. and Lormand, M. and Losurdo, G. and Lough, J. D. and Lousto, C. O. and Lovelace, G. and L\"uck, H. and Lumaca, D. and Lundgren, A. P. and Lynch, R. and Ma, Y. and Macas, R. and Macfoy, S. and Machenschalk, B. and MacInnis, M. and Macleod, D. M. and Maga\~na Hernandez, I. and Maga\~na-Sandoval, F. and Maga\~na Zertuche, L. and Magee, R. M. and Majorana, E. and Maksimovic, I. and Man, N. and Mandic, V. and Mangano, V. and Mansell, G. L. and Manske, M. and Mantovani, M. and Marchesoni, F. and Marion, F. and M\'arka, S. and M\'arka, Z. and Markakis, C. and Markosyan, A. S. and Markowitz, A. and Maros, E. and Marquina, A. and Marsh, P. and Martelli, F. and Martellini, L. and Martin, I. W. and Martin, R. M. and Martynov, D. V. and Marx, J. N. and Mason, K. and Massera, E. and Masserot, A. and Massinger, T. J. and Masso-Reid, M. and Mastrogiovanni, S. and Matas, A. and Matichard, F. and Matone, L. and Mavalvala, N. and Mazumder, N. and McCarthy, R. and McClelland, D. E. and McCormick, S. and McCuller, L. and McGuire, S. C. and McIntyre, G. and McIver, J. and McManus, D. J. and McNeill, L. and McRae, T. and McWilliams, S. T. and Meacher, D. and Meadors, G. D. and Mehmet, M. and Meidam, J. and Mejuto-Villa, E. and Melatos, A. and Mendell, G. and Mercer, R. A. and Merilh, E. L. and Merzougui, M. and Meshkov, S. and Messenger, C. and Messick, C. and Metzdorff, R. and Meyers, P. M. and Miao, H. and Michel, C. and Middleton, H. and Mikhailov, E. E. and Milano, L. and Miller, A. L. and Miller, B. B. and Miller, J. and Millhouse, M. and Milovich-Goff, M. C. and Minazzoli, O. and Minenkov, Y. and Ming, J. and Mishra, C. and Mitra, S. and Mitrofanov, V. P. and Mitselmakher, G. and Mittleman, R. and Moffa, D. and Moggi, A. and Mogushi, K. and Mohan, M. and Mohapatra, S. R. P. and Molina, I. and Montani, M. and Moore, C. J. and Moraru, D. and Moreno, G. and Morisaki, S. and Morriss, S. R. and Mours, B. and Mow-Lowry, C. M. and Mueller, G. and Muir, A. W. and Mukherjee, Arunava and Mukherjee, D. and Mukherjee, S. and Mukund, N. and Mullavey, A. and Munch, J. and Mu\~niz, E. A. and Muratore, M. and Murray, P. G. and Nagar, A. and Napier, K. and Nardecchia, I. and Naticchioni, L. and Nayak, R. K. and Neilson, J. and Nelemans, G. and Nelson, T. J. N. and Nery, M. and Neunzert, A. and Nevin, L. and Newport, J. M. and Newton, G. and Ng, K. K. Y. and Nguyen, P. and Nguyen, T. T. and Nichols, D. and Nielsen, A. B. and Nissanke, S. and Nitz, A. and Noack, A. and Nocera, F. and Nolting, D. and North, C. and Nuttall, L. K. and Oberling, J. and O'Dea, G. D. and Ogin, G. H. and Oh, J. J. and Oh, S. H. and Ohme, F. and Okada, M. A. and Oliver, M. and Oppermann, P. and Oram, Richard J. and O'Reilly, B. and Ormiston, R. and Ortega, L. F. and O'Shaughnessy, R. and Ossokine, S. and Ottaway, D. J. and Overmier, H. and Owen, B. J. and Pace, A. E. and Page, J. and Page, M. A. and Pai, A. and Pai, S. A. and Palamos, J. R. and Palashov, O. and Palomba, C. and Pal-Singh, A. and Pan, Howard and Pan, Huang-Wei and Pang, B. and Pang, P. T. H. and Pankow, C. and Pannarale, F. and Pant, B. C. and Paoletti, F. and Paoli, A. and Papa, M. A. and Parida, A. and Parker, W. and Pascucci, D. and Pasqualetti, A. and Passaquieti, R. and Passuello, D. and Patil, M. and Patricelli, B. and Pearlstone, B. L. and Pedraza, M. and Pedurand, R. and Pekowsky, L. and Pele, A. and Penn, S. and Perez, C. J. and Perreca, A. and Perri, L. M. and Pfeiffer, H. P. and Phelps, M. and Piccinni, O. J. and Pichot, M. and Piergiovanni, F. and Pierro, V. and Pillant, G. and Pinard, L. and Pinto, I. M. and Pirello, M. and Pitkin, M. and Poe, M. and Poggiani, R. and Popolizio, P. and Porter, E. K. and Post, A. and Powell, J. and Prasad, J. and Pratt, J. W. W. and Pratten, G. and Predoi, V. and Prestegard, T. and Prijatelj, M. and Principe, M. and Privitera, S. and Prix, R. and Prodi, G. A. and Prokhorov, L. G. and Puncken, O. and Punturo, M. and Puppo, P. and P\"urrer, M. and Qi, H. and Quetschke, V. and Quintero, E. A. and Quitzow-James, R. and Raab, F. J. and Rabeling, D. S. and Radkins, H. and Raffai, P. and Raja, S. and Rajan, C. and Rajbhandari, B. and Rakhmanov, M. and Ramirez, K. E. and Ramos-Buades, A. and Rapagnani, P. and Raymond, V. and Razzano, M. and Read, J. and Regimbau, T. and Rei, L. and Reid, S. and Reitze, D. H. and Ren, W. and Reyes, S. D. and Ricci, F. and Ricker, P. M. and Rieger, S. and Riles, K. and Rizzo, M. and Robertson, N. A. and Robie, R. and Robinet, F. and Rocchi, A. and Rolland, L. and Rollins, J. G. and Roma, V. J. and Romano, J. D. and Romano, R. and Romel, C. L. and Romie, J. H. and Rosi\ifmmode \acute{n}\else \'{n}\fi{}ska, D. and Ross, M. P. and Rowan, S. and R\"udiger, A. and Ruggi, P. and Rutins, G. and Ryan, K. and Sachdev, S. and Sadecki, T. and Sadeghian, L. and Sakellariadou, M. and Salconi, L. and Saleem, M. and Salemi, F. and Samajdar, A. and Sammut, L. and Sampson, L. M. and Sanchez, E. J. and Sanchez, L. E. and Sanchis-Gual, N. and Sandberg, V. and Sanders, J. R. and Sassolas, B. and Sathyaprakash, B. S. and Saulson, P. R. and Sauter, O. and Savage, R. L. and Sawadsky, A. and Schale, P. and Scheel, M. and Scheuer, J. and Schmidt, J. and Schmidt, P. and Schnabel, R. and Schofield, R. M. S. and Sch\"onbeck, A. and Schreiber, E. and Schuette, D. and Schulte, B. W. and Schutz, B. F. and Schwalbe, S. G. and Scott, J. and Scott, S. M. and Seidel, E. and Sellers, D. and Sengupta, A. S. and Sentenac, D. and Sequino, V. and Sergeev, A. and Shaddock, D. A. and Shaffer, T. J. and Shah, A. A. and Shahriar, M. S. and Shaner, M. B. and Shao, L. and Shapiro, B. and Shawhan, P. and Sheperd, A. and Shoemaker, D. H. and Shoemaker, D. M. and Siellez, K. and Siemens, X. and Sieniawska, M. and Sigg, D. and Silva, A. D. and Singer, L. P. and Singh, A. and Singhal, A. and Sintes, A. M. and Slagmolen, B. J. J. and Smith, B. and Smith, J. R. and Smith, R. J. E. and Somala, S. and Son, E. J. and Sonnenberg, J. A. and Sorazu, B. and Sorrentino, F. and Souradeep, T. and Spencer, A. P. and Srivastava, A. K. and Staats, K. and Staley, A. and Steinke, M. and Steinlechner, J. and Steinlechner, S. and Steinmeyer, D. and Stevenson, S. P. and Stone, R. and Stops, D. J. and Strain, K. A. and Stratta, G. and Strigin, S. E. and Strunk, A. and Sturani, R. and Stuver, A. L. and Summerscales, T. Z. and Sun, L. and Sunil, S. and Suresh, J. and Sutton, P. J. and Swinkels, B. L. and Szczepa\ifmmode \acute{n}\else \'{n}\fi{}czyk, M. J. and Tacca, M. and Tait, S. C. and Talbot, C. and Talukder, D. and Tanner, D. B. and T\'apai, M. and Taracchini, A. and Tasson, J. D. and Taylor, J. A. and Taylor, R. and Tewari, S. V. and Theeg, T. and Thies, F. and Thomas, E. G. and Thomas, M. and Thomas, P. and Thorne, K. A. and Thorne, K. S. and Thrane, E. and Tiwari, S. and Tiwari, V. and Tokmakov, K. V. and Toland, K. and Tonelli, M. and Tornasi, Z. and Torres-Forn\'e, A. and Torrie, C. I. and T\"oyr\"a, D. and Travasso, F. and Traylor, G. and Trinastic, J. and Tringali, M. C. and Trozzo, L. and Tsang, K. W. and Tse, M. and Tso, R. and Tsukada, L. and Tsuna, D. and Tuyenbayev, D. and Ueno, K. and Ugolini, D. and Unnikrishnan, C. S. and Urban, A. L. and Usman, S. A. and Vahlbruch, H. and Vajente, G. and Valdes, G. and Vallisneri, M. and van Bakel, N. and van Beuzekom, M. and van den Brand, J. F. J. and Van Den Broeck, C. and Vander-Hyde, D. C. and van der Schaaf, L. and van Heijningen, J. V. and van Veggel, A. A. and Vardaro, M. and Varma, V. and Vass, S. and Vas\'uth, M. and Vecchio, A. and Vedovato, G. and Veitch, J. and Veitch, P. J. and Venkateswara, K. and Venugopalan, G. and Verkindt, D. and Vetrano, F. and Vicer\'e, A. and Viets, A. D. and Vinciguerra, S. and Vine, D. J. and Vinet, J.-Y. and Vitale, S. and Vo, T. and Vocca, H. and Vorvick, C. and Vyatchanin, S. P. and Wade, A. R. and Wade, L. E. and Wade, M. and Walet, R. and Walker, M. and Wallace, L. and Walsh, S. and Wang, G. and Wang, H. and Wang, J. Z. and Wang, W. H. and Wang, Y. F. and Ward, R. L. and Warner, J. and Was, M. and Watchi, J. and Weaver, B. and Wei, L.-W. and Weinert, M. and Weinstein, A. J. and Weiss, R. and Wen, L. and Wessel, E. K. and We\ss{}els, P. and Westerweck, J. and Westphal, T. and Wette, K. and Whelan, J. T. and Whitcomb, S. E. and Whiting, B. F. and Whittle, C. and Wilken, D. and Williams, D. and Williams, R. D. and Williamson, A. R. and Willis, J. L. and Willke, B. and Wimmer, M. H. and Winkler, W. and Wipf, C. C. and Wittel, H. and Woan, G. and Woehler, J. and Wofford, J. and Wong, K. W. K. and Worden, J. and Wright, J. L. and Wu, D. S. and Wysocki, D. M. and Xiao, S. and Yamamoto, H. and Yancey, C. C. and Yang, L. and Yap, M. J. and Yazback, M. and Yu, Hang and Yu, Haocun and Yvert, M. and Zadro\ifmmode \dot{z}\else \.{z}\fi{}ny, A. and Zanolin, M. and Zelenova, T. and Zendri, J.-P. and Zevin, M. and Zhang, L. and Zhang, M. and Zhang, T. and Zhang, Y.-H. and Zhao, C. and Zhou, M. and Zhou, Z. and Zhu, S. J. and Zhu, X. J. and Zimmerman, A. B. and Zucker, M. E. and Zweizig, J.},
  collaboration = {LIGO Scientific Collaboration and Virgo Collaboration},
  journal = {Phys. Rev. Lett.},
  volume = {119},
  issue = {16},
  pages = {161101},
  numpages = {18},
  year = {2017},
  month = {Oct},
  publisher = {American Physical Society},
  doi = {10.1103/PhysRevLett.119.161101},
  url = {https://link.aps.org/doi/10.1103/PhysRevLett.119.161101}
}

@ARTICLE{abb17b,
       author = {Abbott, B.~P. and Abbott, R. and {Abbott}, T.~D. and {Acernese}, F. and {Ackley}, K. and {Adams}, C. and {Adams}, T. and {Addesso}, P. and {Adhikari}, R.~X. and {Adya}, V.~B. and {Affeldt}, C. and {Afrough}, M. and {Agarwal}, B. and {Agathos}, M. and {Agatsuma}, K. and {Aggarwal}, N. and {Aguiar}, O.~D. and {Aiello}, L. and {Ain}, A. and {Ajith}, P. and {Allen}, B. and {Allen}, G. and {Allocca}, A. and {Altin}, P.~A. and {Amato}, A. and {Ananyeva}, A. and {Anderson}, S.~B. and {Anderson}, W.~G. and {Angelova}, S.~V. and {Antier}, S. and {Appert}, S. and {Arai}, K. and {Araya}, M.~C. and {Areeda}, J.~S. and {Arnaud}, N. and {Arun}, K.~G. and {Ascenzi}, S. and {Ashton}, G. and {Ast}, M. and {Aston}, S.~M. and {Astone}, P. and {Atallah}, D.~V. and {Aufmuth}, P. and {Aulbert}, C. and {AultONeal}, K. and {Austin}, C. and {Avila-Alvarez}, A. and {Babak}, S. and {Bacon}, P. and {Bader}, M.~K.~M. and {Bae}, S. and {Baker}, P.~T. and {Baldaccini}, F. and {Ballardin}, G. and {Ballmer}, S.~W. and {Banagiri}, S. and {Barayoga}, J.~C. and {Barclay}, S.~E. and {Barish}, B.~C. and {Barker}, D. and {Barkett}, K. and {Barone}, F. and {Barr}, B. and {Barsotti}, L. and {Barsuglia}, M. and {Barta}, D. and {Barthelmy}, S.~D. and {Bartlett}, J. and {Bartos}, I. and {Bassiri}, R. and {Basti}, A. and {Batch}, J.~C. and {Bawaj}, M. and {Bayley}, J.~C. and {Bazzan}, M. and {B{\'e}csy}, B. and {Beer}, C. and {Bejger}, M. and {Belahcene}, I. and {Bell}, A.~S. and {Berger}, B.~K. and {Bergmann}, G. and {Bero}, J.~J. and {Berry}, C.~P.~L. and {Bersanetti}, D. and {Bertolini}, A. and {Betzwieser}, J. and {Bhagwat}, S. and {Bhandare}, R. and {Bilenko}, I.~A. and {Billingsley}, G. and {Billman}, C.~R. and {Birch}, J. and {Birney}, R. and {Birnholtz}, O. and {Biscans}, S. and {Biscoveanu}, S. and {Bisht}, A. and {Bitossi}, M. and {Biwer}, C. and {Bizouard}, M.~A. and {Blackburn}, J.~K. and {Blackman}, J. and {Blair}, C.~D. and {Blair}, D.~G. and {Blair}, R.~M. and {Bloemen}, S. and {Bock}, O. and {Bode}, N. and {Boer}, M. and {Bogaert}, G. and {Bohe}, A. and {Bondu}, F. and {Bonilla}, E. and {Bonnand}, R. and {Boom}, B.~A. and {Bork}, R. and {Boschi}, V. and {Bose}, S. and {Bossie}, K. and {Bouffanais}, Y. and {Bozzi}, A. and {Bradaschia}, C. and {Brady}, P.~R. and {Branchesi}, M. and {Brau}, J.~E. and {Briant}, T. and {Brillet}, A. and {Brinkmann}, M. and {Brisson}, V. and {Brockill}, P. and {Broida}, J.~E. and {Brooks}, A.~F. and {Brown}, D.~A. and {Brown}, D.~D. and {Brunett}, S. and {Buchanan}, C.~C. and {Buikema}, A. and {Bulik}, T. and {Bulten}, H.~J. and {Buonanno}, A. and {Buskulic}, D. and {Buy}, C. and {Byer}, R.~L. and {Cabero}, M. and {Cadonati}, L. and {Cagnoli}, G. and {Cahillane}, C. and {Calder{\'o}n Bustillo}, J. and {Callister}, T.~A. and {Calloni}, E. and {Camp}, J.~B. and {Canepa}, M. and {Canizares}, P. and {Cannon}, K.~C. and {Cao}, H. and {Cao}, J. and {Capano}, C.~D. and {Capocasa}, E. and {Carbognani}, F. and {Caride}, S. and {Carney}, M.~F. and {Casanueva Diaz}, J. and {Casentini}, C. and {Caudill}, S. and {Cavagli{\`a}}, M. and {Cavalier}, F. and {Cavalieri}, R. and {Cella}, G. and {Cepeda}, C.~B. and {Cerd{\'a}-Dur{\'a}n}, P. and {Cerretani}, G. and {Cesarini}, E. and {Chamberlin}, S.~J. and {Chan}, M. and {Chao}, S. and {Charlton}, P. and {Chase}, E. and {Chassande-Mottin}, E. and {Chatterjee}, D. and {Chatziioannou}, K. and {Cheeseboro}, B.~D. and {Chen}, H.~Y. and {Chen}, X. and {Chen}, Y. and {Cheng}, H.-P. and {Chia}, H. and {Chincarini}, A. and {Chiummo}, A. and {Chmiel}, T. and {Cho}, H.~S. and {Cho}, M. and {Chow}, J.~H. and {Christensen}, N. and {Chu}, Q. and {Chua}, A.~J.~K. and {Chua}, S. and {Chung}, A.~K.~W. and {Chung}, S. and {Ciani}, G.},
        title = "{Multi-messenger Observations of a Binary Neutron Star Merger}",
      journal = {\apjl},
         year = 2017,
        month = oct,
       volume = {848},
       number = {2},
          eid = {L12},
        pages = {L12},
          doi = {10.3847/2041-8213/aa91c9},
archivePrefix = {arXiv},
       eprint = {1710.05833},
 primaryClass = {astro-ph.HE},
       adsurl = {https://ui.adsabs.harvard.edu/abs/2017ApJ...848L..12A}
}

@article{abb17GRB,
doi = {10.3847/2041-8213/aa920c},
url = {https://doi.org/10.3847/2041-8213/aa920c},
year = {2017},
month = {oct},
publisher = {The American Astronomical Society},
volume = {848},
number = {2},
pages = {L13},
author = {Abbott, B. P. and Abbott, R. and Abbott, T. D. and Acernese, F. and Ackley, K. and Adams, C. and Adams, T. and Addesso, P. and Adhikari, R. X. and Adya, V. B. and Affeldt, C. and Afrough, M. and Agarwal, B. and Agathos, M. and Agatsuma, K. and Aggarwal, N. and Aguiar, O. D. and Aiello, L. and Ain, A. and Ajith, P. and Allen, B. and Allen, G. and Allocca, A. and Aloy, M. A. and Altin, P. A. and Amato, A. and Ananyeva, A. and Anderson, S. B. and Anderson, W. G. and Angelova, S. V. and Antier, S. and Appert, S. and Arai, K. and Araya, M. C. and Areeda, J. S. and Arnaud, N. and Arun, K. G. and Ascenzi, S. and Ashton, G. and Ast, M. and Aston, S. M. and Astone, P. and Atallah, D. V. and Aufmuth, P. and Aulbert, C. and AultONeal, K. and Austin, C. and Avila-Alvarez, A. and Babak, S. and Bacon, P. and Bader, M. K. M. and Bae, S. and Baker, P. T. and Baldaccini, F. and Ballardin, G. and Ballmer, S. W. and Banagiri, S. and Barayoga, J. C. and Barclay, S. E. and Barish, B. C. and Barker, D. and Barkett, K. and Barone, F. and Barr, B. and Barsotti, L. and Barsuglia, M. and Barta, D. and Bartlett, J. and Bartos, I. and Bassiri, R. and Basti, A. and Batch, J. C. and Bawaj, M. and Bayley, J. C. and Bazzan, M. and Bécsy, B. and Beer, C. and Bejger, M. and Belahcene, I. and Bell, A. S. and Berger, B. K. and Bergmann, G. and Bero, J. J. and Berry, C. P. L. and Bersanetti, D. and Bertolini, A. and Betzwieser, J. and Bhagwat, S. and Bhandare, R. and Bilenko, I. A. and Billingsley, G. and Billman, C. R. and Birch, J. and Birney, I. A. and Birnholtz, O. and Biscans, S. and Biscoveanu, S. and Bisht, A. and Bitossi, M. and Biwer, C. and Bizouard, M. A. and Blackburn, J. K. and Blackman, J. and Blair, C. D. and Blair, D. G. and Blair, R. M. and Bloemen, S. and Bock, O. and Bode, N. and Boer, M. and Bogaert, G. and Bohe, A. and Bondu, F. and Bonilla, E. and Bonnand, R. and Boom, B. A. and Bork, R. and Boschi, V. and Bose, S. and Bossie, K. and Bouffanais, Y. and Bozzi, A. and Bradaschia, C. and Brady, P. R. and Branchesi, M. and Brau, J. E. and Briant, T. and Brillet, A. and Brinkmann, M. and Brisson, V. and Brockill, P. and Broida, J. E. and Brooks, A. F. and Brown, D. A. and Brown, D. D. and Brunett, S. and Buchanan, C. C. and Buikema, A. and Bulik, T. and Bulten, H. J. and Buonanno, A. and Buskulic, D. and Buy, C. and Byer, R. L. and Cabero, M. and Cadonati, L. and Cagnoli, G. and Cahillane, C. and Calderón Bustillo, J. and Callister, T. A. and Calloni, E. and Camp, J. B. and Canepa, M. and Canizares, P. and Cannon, K. C. and Cao, H. and Cao, J. and Capano, C. D. and Capocasa, E. and Carbognani, F. and Caride, S. and Carney, M. F. and Diaz, J. Casanueva and Casentini, C. and Caudill, S. and Cavaglià, M. and Cavalier, F. and Cavalieri, R. and Cella, G. and Cepeda, C. B. and Cerdá-Durán, P. and Cerretani, G. and Cesarini, E. and Chamberlin, S. J. and Chan, M. and Chao, S. and Charlton, P. and Chase, E. and Chassande-Mottin, E. and Chatterjee, D. and Chatziioannou, K. and Cheeseboro, B. D. and Chen, H. Y. and Chen, X. and Chen, Y. and Cheng, H.-P. and Chia, H. and Chincarini, A. and Chiummo, A. and Chmiel, T. and Cho, H. S. and Cho, M. and Chow, J. H. and Christensen, N. and Chu, Q. and Chua, A. J. K. and Chua, S. and Chung, A. K. W. and Chung, S. and Ciani, G. and Ciolfi, R. and Cirelli, C. E. and Cirone, A. and Clara, F. and Clark, J. A. and Clearwater, P. and Cleva, F. and Cocchieri, C. and Coccia, E. and Cohadon, P.-F. and Cohen, D. and Colla, A. and Collette, C. G. and Cominsky, L. R. and Constancio Jr., M. and Conti, L. and Cooper, S. J. and Corban, P. and Corbitt, T. R. and Cordero-Carrión, I. and Corley, K. R. and Cornish, N. and Corsi, A. and Cortese, S. and Costa, C. A. and Coughlin, M. W. and Coughlin, S. B. and Coulon, J.-P. and Countryman, S. T. and Couvares, P. and Covas, P. B. and Cowan, E. E. and Coward, D. M. and Cowart, M. J. and Coyne, D. C. and Coyne, R. and Creighton, J. D. E. and Creighton, T. D. and Cripe, J. and Crowder, S. G. and Cullen, T. J. and Cumming, A. and Cunningham, L. and Cuoco, E. and Canton, T. Dal and Dálya, G. and Danilishin, S. L. and D’Antonio, S. and Danzmann, K. and Dasgupta, A. and Costa, C. F. Da Silva and Dattilo, V. and Dave, I. and Davier, M. and Davis, D. and Daw, E. J. and Day, B. and De, S. and DeBra, D. and Degallaix, J. and Laurentis, M. De and Deléglise, S. and Pozzo, W. Del and Demos, N. and Denker, T. and Dent, T. and Pietri, R. De and Dergachev, V. and Rosa, R. De and DeRosa, R. T. and Rossi, C. De and DeSalvo, R. and Varona, O. de and Devenson, J. and Dhurandhar, S. and Díaz, M. C. and Fiore, L. Di and Giovanni, M. Di and Girolamo, T. Di and Lieto, A. Di and Pace, S. Di and Palma, I. Di and Renzo, F. Di and Doctor, Z. and Dolique, V. and Donovan, F. and Dooley, K. L. and Doravari, S. and Dorrington, I. and Douglas, R. and Dovale Álvarez, M. and Downes, T. P. and Drago, M. and Dreissigacker, C. and Driggers, J. C. and Du, Z. and Ducrot, M. and Dupej, P. and Dwyer, S. E. and Edo, T. B. and Edwards, M. C. and Effler, A. and Eggenstein, H.-B. and Ehrens, P. and Eichholz, J. and Eikenberry, S. S. and Eisenstein, R. A. and Essick, R. C. and Estevez, D. and Etienne, Z. B. and Etzel, T. and Evans, M. and Evans, T. M. and Factourovich, M. and Fafone, V. and Fair, H. and Fairhurst, S. and Fan, X. and Farinon, S. and Farr, B. and Farr, W. M. and Fauchon-Jones, E. J. and Favata, M. and Fays, M. and Fee, C. and Fehrmann, H. and Feicht, J. and Fejer, M. M. and Fernandez-Galiana, A. and Ferrante, I. and Ferreira, E. C. and Ferrini, F. and Fidecaro, F. and Finstad, D. and Fiori, I. and Fiorucci, D. and Fishbach, M. and Fisher, R. P. and Fitz-Axen, M. and Flaminio, R. and Fletcher, M. and Fong, H. and Font, J. A. and Forsyth, P. W. F. and Forsyth, S. S. and Fournier, J.-D. and Frasca, S. and Frasconi, F. and Frei, Z. and Freise, A. and Frey, R. and Frey, V. and Fries, E. M. and Fritschel, P. and Frolov, V. V. and Fulda, P. and Fyffe, M. and Gabbard, H. and Gadre, B. U. and Gaebel, S. M. and Gair, J. R. and Gammaitoni, L. and Ganija, M. R. and Gaonkar, S. G. and Garcia-Quiros, C. and Garufi, F. and Gateley, B. and Gaudio, S. and Gaur, G. and Gayathri, V. and Gehrels, N. and Gemme, G. and Genin, E. and Gennai, A. and George, D. and George, J. and Gergely, L. and Germain, V. and Ghonge, S. and Ghosh, Abhirup and Ghosh, Archisman and Ghosh, S. and Giaime, J. A. and Giardina, K. D. and Giazotto, A. and Gill, K. and Glover, L. and Goetz, E. and Goetz, R. and Gomes, S. and Goncharov, B. and González, G. and Castro, J. M. Gonzalez and Gopakumar, A. and Gorodetsky, M. L. and Gossan, S. E. and Gosselin, M. and Gouaty, R. and Grado, A. and Graef, C. and Granata, M. and Grant, A. and Gras, S. and Gray, C. and Greco, G. and Green, A. C. and Gretarsson, E. M. and Groot, P. and Grote, H. and Grunewald, S. and Gruning, P. and Guidi, G. M. and Guo, X. and Gupta, A. and Gupta, M. K. and Gushwa, K. E. and Gustafson, E. K. and Gustafson, R. and Halim, O. and Hall, B. R. and Hall, E. D. and Hamilton, E. Z. and Hammond, G. and Haney, M. and Hanke, M. M. and Hanks, J. and Hanna, C. and Hannam, M. D. and Hannuksela, O. A. and Hanson, J. and Hardwick, T. and Harms, J. and Harry, G. M. and Harry, I. W. and Hart, M. J. and Haster, C.-J. and Haughian, K. and Healy, J. and Heidmann, A. and Heintze, M. C. and Heitmann, H. and Hello, P. and Hemming, G. and Hendry, M. and Heng, I. S. and Hennig, J. and Heptonstall, A. W. and Heurs, M. and Hild, S. and Hinderer, T. and Hoak, D. and Hofman, D. and Holt, K. and Holz, D. E. and Hopkins, P. and Horst, C. and Hough, J. and Houston, E. A. and Howell, E. J. and Hreibi, A. and Hu, Y. M. and Huerta, E. A. and Huet, D. and Hughey, B. and Husa, S. and Huttner, S. H. and Huynh-Dinh, T. and Indik, N. and Inta, R. and Intini, G. and Isa, H. N. and Isac, J.-M. and Isi, M. and Iyer, B. R. and Izumi, K. and Jacqmin, T. and Jani, K. and Jaranowski, P. and Jawahar, S. and Jiménez-Forteza, F. and Johnson, W. W. and Johnson-McDaniel, N. K. and Jones, D. I. and Jones, R. and Jonker, R. J. G. and Ju, L. and Junker, J. and Kalaghatgi, C. V. and Kalogera, V. and Kamai, B. and Kandhasamy, S. and Kang, G. and Kanner, J. B. and Kapadia, S. J. and Karki, S. and Karvinen, K. S. and Kasprzack, M. and Kastaun, W. and Katolik, M. and Katsavounidis, E. and Katzman, W. and Kaufer, S. and Kawabe, K. and Kéfélian, F. and Keitel, D. and Kemball, A. J. and Kennedy, R. and Kent, C. and Key, J. S. and Khalili, F. Y. and Khan, I. and Khan, S. and Khan, Z. and Khazanov, E. A. and Kijbunchoo, N. and Kim, Chunglee and Kim, J. C. and Kim, K. and Kim, W. and Kim, W. S. and Kim, Y.-M. and Kimbrell, S. J. and King, E. J. and King, P. J. and Kinley-Hanlon, M. and Kirchhoff, R. and Kissel, J. S. and Kleybolte, L. and Klimenko, S. and Knowles, T. D. and Koch, P. and Koehlenbeck, S. M. and Koley, S. and Kondrashov, V. and Kontos, A. and Korobko, M. and Korth, W. Z. and Kowalska, I. and Kozak, D. B. and Krämer, C. and Kringel, V. and Krishnan, B. and Królak, A. and Kuehn, G. and Kumar, P. and Kumar, R. and Kumar, S. and Kuo, L. and Kutynia, A. and Kwang, S. and Lackey, B. D. and Lai, K. H. and Landry, M. and Lang, R. N. and Lange, J. and Lantz, B. and Lanza, R. K. and Lartaux-Vollard, A. and Lasky, P. D. and Laxen, M. and Lazzarini, A. and Lazzaro, C. and Leaci, P. and Leavey, S. and Lee, C. H. and Lee, H. K. and Lee, H. M. and Lee, H. W. and Lee, K. and Lehmann, J. and Lenon, A. and Leonardi, M. and Leroy, N. and Letendre, N. and Levin, Y. and Li, T. G. F. and Linker, S. D. and Littenberg, T. B. and Liu, J. and Lo, R. K. L. and Lockerbie, N. A. and London, L. T. and Lord, J. E. and Lorenzini, M. and Loriette, V. and Lormand, M. and Losurdo, G. and Lough, J. D. and Lousto, C. O. and Lovelace, G. and Lück, H. and Lumaca, D. and Lundgren, A. P. and Lynch, R. and Ma, Y. and Macas, R. and Macfoy, S. and Machenschalk, B. and MacInnis, M. and Macleod, D. M. and Magaña Hernandez, I. and Magaña-Sandoval, F. and Magaña Zertuche, L. and Magee, R. M. and Majorana, E. and Maksimovic, I. and Man, N. and Mandic, V. and Mangano, V. and Mansell, G. L. and Manske, M. and Mantovani, M. and Marchesoni, F. and Marion, F. and Márka, S. and Márka, Z. and Markakis, C. and Markosyan, A. S. and Markowitz, A. and Maros, E. and Marquina, A. and Martelli, F. and Martellini, L. and Martin, I. W. and Martin, R. M. and Martynov, D. V. and Mason, K. and Massera, E. and Masserot, A. and Massinger, T. J. and Masso-Reid, M. and Mastrogiovanni, S. and Matas, A. and Matichard, F. and Matone, L. and Mavalvala, N. and Mazumder, N. and McCarthy, R. and McClelland, D. E. and McCormick, S. and McCuller, L. and McGuire, S. C. and McIntyre, G. and McIver, J. and McManus, D. J. and McNeill, L. and McRae, T. and McWilliams, S. T. and Meacher, D. and Meadors, G. D. and Mehmet, M. and Meidam, J. and Mejuto-Villa, E. and Melatos, A. and Mendell, G. and Mercer, R. A. and Merilh, E. L. and Merzougui, M. and Meshkov, S. and Messenger, C. and Messick, C. and Metzdorff, R. and Meyers, P. M. and Miao, H. and Michel, C. and Middleton, H. and Mikhailov, E. E. and Milano, L. and Miller, A. L. and Miller, B. B. and Miller, J. and Millhouse, M. and Milovich-Goff, M. C. and Minazzoli, O. and Minenkov, Y. and Ming, J. and Mishra, C. and Mitra, S. and Mitrofanov, V. P. and Mitselmakher, G. and Mittleman, R. and Moffa, D. and Moggi, A. and Mogushi, K. and Mohan, M. and Mohapatra, S. R. P. and Montani, M. and Moore, C. J. and Moraru, D. and Moreno, G. and Morriss, S. R. and Mours, B. and Mow-Lowry, C. M. and Mueller, G. and Muir, A. W. and Mukherjee, Arunava and Mukherjee, D. and Mukherjee, S. and Mukund, N. and Mullavey, A. and Munch, J. and Muñiz, E. A. and Muratore, M. and Murray, P. G. and Napier, K. and Nardecchia, I. and Naticchioni, L. and Nayak, R. K. and Neilson, J. and Nelemans, G. and Nelson, T. J. N. and Nery, M. and Neunzert, A. and Nevin, L. and Newport, J. M. and Newton, G. and Ng, K. K. Y. and Nguyen, T. T. and Nichols, D. and Nielsen, A. B. and Nissanke, S. and Nitz, A. and Noack, A. and Nocera, F. and Nolting, D. and North, C. and Nuttall, L. K. and Oberling, J. and O’Dea, G. D. and Ogin, G. H. and Oh, J. J. and Oh, S. H. and Ohme, F. and Okada, M. A. and Oliver, M. and Oppermann, P. and Oram, Richard J. and O’Reilly, B. and Ormiston, R. and Ortega, L. F. and O’Shaughnessy, R. and Ossokine, S. and Ottaway, D. J. and Overmier, H. and Owen, B. J. and Pace, A. E. and Page, J. and Page, M. A. and Pai, A. and Pai, S. A. and Palamos, J. R. and Palashov, O. and Palomba, C. and Pal-Singh, A. and Pan, Howard and Pan, Huang-Wei and Pang, B. and Pang, P. T. H. and Pankow, C. and Pannarale, F. and Pant, B. C. and Paoletti, F. and Paoli, A. and Papa, M. A. and Parida, A. and Parker, W. and Pascucci, D. and Pasqualetti, A. and Passaquieti, R. and Passuello, D. and Patil, M. and Patricelli, B. and Pearlstone, B. L. and Pedraza, M. and Pedurand, R. and Pekowsky, L. and Pele, A. and Penn, S. and Perez, C. J. and Perreca, A. and Perri, L. M. and Pfeiffer, H. P. and Phelps, M. and Piccinni, O. J. and Pichot, M. and Piergiovanni, F. and Pierro, V. and Pillant, G. and Pinard, L. and Pinto, I. M. and Pirello, M. and Pitkin, M. and Poe, M. and Poggiani, R. and Popolizio, P. and Porter, E. K. and Post, A. and Powell, J. and Prasad, J. and Pratt, J. W. W. and Pratten, G. and Predoi, V. and Prestegard, T. and Prijatelj, M. and Principe, M. and Privitera, S. and Prodi, G. A. and Prokhorov, L. G. and Puncken, O. and Punturo, M. and Puppo, P. and Pürrer, M. and Qi, H. and Quetschke, V. and Quintero, E. A. and Quitzow-James, R. and Raab, F. J. and Rabeling, D. S. and Radkins, H. and Raffai, P. and Raja, S. and Rajan, C. and Rajbhandari, B. and Rakhmanov, M. and Ramirez, K. E. and Ramos-Buades, A. and Rapagnani, P. and Raymond, V. and Razzano, M. and Read, J. and Regimbau, T. and Rei, L. and Reid, S. and Reitze, D. H. and Ren, W. and Reyes, S. D. and Ricci, F. and Ricker, P. M. and Rieger, S. and Riles, K. and Rizzo, M. and Robertson, N. A. and Robie, R. and Robinet, F. and Rocchi, A. and Rolland, L. and Rollins, J. G. and Roma, V. J. and Romano, R. and Romel, C. L. and Romie, J. H. and Rosińska, D. and Ross, M. P. and Rowan, S. and Rüdiger, A. and Ruggi, P. and Rutins, G. and Ryan, K. and Sachdev, S. and Sadecki, T. and Sadeghian, L. and Sakellariadou, M. and Salconi, L. and Saleem, M. and Salemi, F. and Samajdar, A. and Sammut, L. and Sampson, L. M. and Sanchez, E. J. and Sanchez, L. E. and Sanchis-Gual, N. and Sandberg, V. and Sanders, J. R. and Sassolas, B. and Sathyaprakash, B. S. and Saulson, P. R. and Sauter, O. and Savage, R. L. and Sawadsky, A. and Schale, P. and Scheel, M. and Scheuer, J. and Schmidt, J. and Schmidt, P. and Schnabel, R. and Schofield, R. M. S. and Schönbeck, A. and Schreiber, E. and Schuette, D. and Schulte, B. W. and Schutz, B. F. and Schwalbe, S. G. and Scott, J. and Scott, S. M. and Seidel, E. and Sellers, D. and Sengupta, A. S. and Sentenac, D. and Sequino, V. and Sergeev, A. and Shaddock, D. A. and Shaffer, T. J. and Shah, A. A. and Shahriar, M. S. and Shaner, M. B. and Shao, L. and Shapiro, B. and Shawhan, P. and Sheperd, A. and Shoemaker, D. H. and Shoemaker, D. M. and Siellez, K. and Siemens, X. and Sieniawska, M. and Sigg, D. and Silva, A. D. and Singer, L. P. and Singh, A. and Singhal, A. and Sintes, A. M. and Slagmolen, B. J. J. and Smith, B. and Smith, J. R. and Smith, R. J. E. and Somala, S. and Son, E. J. and Sonnenberg, J. A. and Sorazu, B. and Sorrentino, F. and Souradeep, T. and Spencer, A. P. and Srivastava, A. K. and Staats, K. and Staley, A. and Steinke, M. and Steinlechner, J. and Steinlechner, S. and Steinmeyer, D. and Stevenson, S. P. and Stone, R. and Stops, D. J. and Strain, K. A. and Stratta, G. and Strigin, S. E. and Strunk, A. and Sturani, R. and Stuver, A. L. and Summerscales, T. Z. and Sun, L. and Sunil, S. and Suresh, J. and Sutton, P. J. and Swinkels, B. L. and Szczepańczyk, M. J. and Tacca, M. and Tait, S. C. and Talbot, C. and Talukder, D. and Tanner, D. B. and Tápai, M. and Taracchini, A. and Tasson, J. D. and Taylor, J. A. and Taylor, R. and Tewari, S. V. and Theeg, T. and Thies, F. and Thomas, E. G. and Thomas, M. and Thomas, P. and Thorne, K. A. and Thorne, K. S. and Thrane, E. and Tiwari, S. and Tiwari, V. and Tokmakov, K. V. and Toland, K. and Tonelli, M. and Tornasi, Z. and Torres-Forné, A. and Torrie, C. I. and Töyrä, D. and Travasso, F. and Traylor, G. and Trinastic, J. and Tringali, M. C. and Trozzo, L. and Tsang, K. W. and Tse, M. and Tso, R. and Tsukada, L. and Tsuna, D. and Tuyenbayev, D. and Ueno, K. and Ugolini, D. and Unnikrishnan, C. S. and Urban, A. L. and Usman, S. A. and Vahlbruch, H. and Vajente, G. and Valdes, G. and Bakel, N. van and Beuzekom, M. van and Brand, J. F. J. van den and Broeck, C. Van Den and Vander-Hyde, D. C. and Schaaf, L. van der and Heijningen, J. V. van and Veggel, A. A. van and Vardaro, M. and Varma, V. and Vass, S. and Vasúth, M. and Vecchio, A. and Vedovato, G. and Veitch, J. and Veitch, P. J. and Venkateswara, K. and Venugopalan, G. and Verkindt, D. and Vetrano, F. and Viceré, A. and Viets, A. D. and Vinciguerra, S. and Vine, D. J. and Vinet, J.-Y. and Vitale, S. and Vo, T. and Vocca, H. and Vorvick, C. and Vyatchanin, S. P. and Wade, A. R. and Wade, L. E. and Wade, M. and Walet, R. and Walker, M. and Wallace, L. and Walsh, S. and Wang, G. and Wang, H. and Wang, J. Z. and Wang, W. H. and Wang, Y. F. and Ward, R. L. and Warner, J. and Was, M. and Watchi, J. and Weaver, B. and Wei, L.-W. and Weinert, M. and Weinstein, A. J. and Weiss, R. and Wen, L. and Wessel, E. K. and Weßels, P. and Westerweck, J. and Westphal, T. and Wette, K. and Whelan, J. T. and Whitcomb, S. E. and Whiting, B. F. and Whittle, C. and Wilken, D. and Williams, D. and Williams, R. D. and Williamson, A. R. and Willis, J. L. and Willke, B. and Wimmer, M. H. and Winkler, W. and Wipf, C. C. and Wittel, H. and Woan, G. and Woehler, J. and Wofford, J. and Wong, K. W. K. and Worden, J. and Wright, J. L. and Wu, D. S. and Wysocki, D. M. and Xiao, S. and Yamamoto, H. and Yancey, C. C. and Yang, L. and Yap, M. J. and Yazback, M. and Yu, Hang and Yu, Haocun and Yvert, M. and Zadrożny, A. and Zanolin, M. and Zelenova, T. and Zendri, J.-P. and Zevin, M. and Zhang, L. and Zhang, M. and Zhang, T. and Zhang, Y.-H. and Zhao, C. and Zhou, M. and Zhou, Z. and Zhu, S. J. and Zhu, X. J. and Zimmerman, A. B. and Zucker, M. E. and Zweizig, J. and (LIGO Scientific Collaboration and Virgo Collaboration) and Burns, E. and Veres, P. and Kocevski, D. and Racusin, J. and Goldstein, A. and Connaughton, V. and Briggs, M. S. and Blackburn, L. and Hamburg, R. and Hui, C. M. and Kienlin, A. von and McEnery, J. and Preece, R. D. and Wilson-Hodge, C. A. and Bissaldi, E. and Cleveland, W. H. and Gibby, M. H. and Giles, M. M. and Kippen, R. M. and McBreen, S. and Meegan, C. A. and Paciesas, W. S. and Poolakkil, S. and Roberts, O. J. and Stanbro, M. and (Fermi Gamma-ray Burst Monitor) and Savchenko, V. and Ferrigno, C. and Kuulkers, E. and Bazzano, A. and Bozzo, E. and Brandt, S. and Chenevez, J. and Courvoisier, T. J.-L. and Diehl, R. and Domingo, A. and Hanlon, L. and Jourdain, E. and Laurent, P. and Lebrun, F. and Lutovinov, A. and Mereghetti, S. and Natalucci, L. and Rodi, J. and Roques, J.-P. and Sunyaev, R. and Ubertini, P. and (INTEGRAL)},
title = {Gravitational Waves and Gamma-Rays from a Binary Neutron Star Merger: GW170817 and GRB 170817A},
journal = {The Astrophysical Journal Letters}
}

@article{abb19,
  title = {GWTC-1: A Gravitational-Wave Transient Catalog of Compact Binary Mergers Observed by LIGO and Virgo during the First and Second Observing Runs},
  author = {Abbott, B. P. and others},
  collaboration = {LIGO Scientific Collaboration and Virgo Collaboration},
  journal = {Phys. Rev. X},
  volume = {9},
  issue = {3},
  pages = {031040},
  numpages = {49},
  year = {2019},
  month = {Sep},
  publisher = {American Physical Society},
  doi = {10.1103/PhysRevX.9.031040},
  url = {https://link.aps.org/doi/10.1103/PhysRevX.9.031040}
}

@article{tan17,
doi = {10.3847/2041-8213/aa90b6},
url = {https://doi.org/10.3847/2041-8213/aa90b6},
year = {2017},
month = {oct},
publisher = {The American Astronomical Society},
volume = {848},
number = {2},
pages = {L27},
author = {Tanvir, N. R. and Levan, A. J. and Gonz{\'a}lez-Fern{\'a}ndez, C. and Korobkin, O. and Mandel, I. and Rosswog, S. and Hjorth, J. and D’Avanzo, P. and Fruchter, A. S. and Fryer, C. L. and Kangas, T. and Milvang-Jensen, B. and Rosetti, S. and Steeghs, D. and Wollaeger, R. T. and Cano, Z. and Copperwheat, C. M. and Covino, S. and D’Elia, V. and de Ugarte Postigo, A. and Evans, P. A. and Even, W. P. and Fairhurst, S. and Jaimes, R. Figuera and Fontes, C. J. and Fujii, Y. I. and Fynbo, J. P. U. and Gompertz, B. P. and Greiner, J. and Hodosan, G. and Irwin, M. J. and Jakobsson, P. and Jørgensen, U. G. and Kann, D. A. and Lyman, J. D. and Malesani, D. and McMahon, R. G. and Melandri, A. and O’Brien, P. T. and Osborne, J. P. and Palazzi, E. and Perley, D. A. and Pian, E. and Piranomonte, S. and Rabus, M. and Rol, E. and Rowlinson, A. and Schulze, S. and Sutton, P. and Thöne, C. C. and Ulaczyk, K. and Watson, D. and Wiersema, K. and Wijers, R. A. M. J.},
title = {The Emergence of a Lanthanide-rich Kilonova Following the Merger of Two Neutron Stars},
journal = {The Astrophysical Journal Letters}
}

@article{poz18,
doi = {10.3847/2041-8213/aaa2f6},
url = {https://doi.org/10.3847/2041-8213/aaa2f6},
year = {2018},
month = {jan},
publisher = {The American Astronomical Society},
volume = {852},
number = {2},
pages = {L30},
author = {Pozanenko, A. S. and Barkov, M. V. and Minaev, P. Yu. and Volnova, A. A. and Mazaeva, E. D. and Moskvitin, A. S. and Krugov, M. A. and Samodurov, V. A. and Loznikov, V. M. and Lyutikov, M.},
title = {GRB 170817A Associated with GW170817: Multi-frequency Observations and Modeling of Prompt Gamma-Ray Emission},
journal = {The Astrophysical Journal Letters}
}

@article{com23,
    author = "Combi, Luciano and Siegel, Daniel M.",
    title = "{Jets from Neutron-Star Merger Remnants and Massive Blue Kilonovae}",
    eprint = "2303.12284",
    archivePrefix = "arXiv",
    primaryClass = "astro-ph.HE",
    doi = "10.1103/PhysRevLett.131.231402",
    journal = "Phys. Rev. Lett.",
    volume = "131",
    number = "23",
    pages = "231402",
    year = "2023"
}

@article{wol21,
doi = {10.3847/1538-4357/ac0d03},
url = {https://doi.org/10.3847/1538-4357/ac0d03},
year = {2021},
month = {aug},
publisher = {The American Astronomical Society},
volume = {918},
number = {1},
pages = {10},
author = {Wollaeger, R. T. and Fryer, C. L. and Chase, E. A. and Fontes, C. J. and Ristic, M. and Hungerford, A. L. and Korobkin, O. and O’Shaughnessy, R. and Herring, A. M.},
title = {A Broad Grid of 2D Kilonova Emission Models},
journal = {The Astrophysical Journal}
}

@article{ban20,
doi = {10.3847/1538-4357/abae61},
url = {https://doi.org/10.3847/1538-4357/abae61},
year = {2020},
month = {sep},
publisher = {The American Astronomical Society},
volume = {901},
number = {1},
pages = {29},
author = {Banerjee, Smaranika and Tanaka, Masaomi and Kawaguchi, Kyohei and Kato, Daiji and Gaigalas, Gediminas},
title = {Simulations of Early Kilonova Emission from Neutron Star Mergers},
journal = {The Astrophysical Journal}
}

@article{ban22,
doi = {10.3847/1538-4357/ac7565},
url = {https://doi.org/10.3847/1538-4357/ac7565},
year = {2022},
month = {jul},
publisher = {The American Astronomical Society},
volume = {934},
number = {2},
pages = {117},
author = {Banerjee, Smaranika and Tanaka, Masaomi and Kato, Daiji and Gaigalas, Gediminas and Kawaguchi, Kyohei and Domoto, Nanae},
title = {Opacity of the Highly Ionized Lanthanides and the Effect on the Early Kilonova},
journal = {The Astrophysical Journal}
}

@ARTICLE{ban24,
       author = {Banerjee, Smaranika and {Tanaka}, Masaomi and {Kato}, Daiji and {Gaigalas}, Gediminas},
        title = "{Diversity of Early Kilonova with the Realistic Opacities of Highly Ionized Heavy Elements}",
      journal = {\apj},
         year = 2024,
        month = jun,
       volume = {968},
       number = {2},
          eid = {64},
        pages = {64},
          doi = {10.3847/1538-4357/ad4029},
archivePrefix = {arXiv},
       eprint = {2304.05810},
 primaryClass = {astro-ph.HE},
       adsurl = {https://ui.adsabs.harvard.edu/abs/2024ApJ...968...64B}
}

@article{lof24,
    author = "Loffredo, E. and others",
    title = "{Prospects for optical detections from binary neutron star mergers with the next-generation multi-messenger observatories}",
    eprint = "2411.02342",
    archivePrefix = "arXiv",
    primaryClass = "astro-ph.HE",
    doi = "10.1051/0004-6361/202452863",
    journal = "Astron. Astrophys.",
    volume = "697",
    pages = "A36",
    year = "2025"
}

@ARTICLE{ai25,
       author = {{Ai}, Shunke and {Gao}, He and {Zhang}, Bing},
        title = "{Engine-fed Kilonovae (Mergernovae). II. Radiation}",
      journal = {\apj},
         year = 2025,
        month = jan,
       volume = {978},
       number = {1},
          eid = {52},
        pages = {52},
          doi = {10.3847/1538-4357/ad93b4},
archivePrefix = {arXiv},
       eprint = {2405.00638},
 primaryClass = {astro-ph.HE},
       adsurl = {https://ui.adsabs.harvard.edu/abs/2025ApJ...978...52A}
}

@article{ale17,
doi = {10.3847/2041-8213/aa905d},
url = {https://doi.org/10.3847/2041-8213/aa905d},
year = {2017},
month = {oct},
publisher = {The American Astronomical Society},
volume = {848},
number = {2},
pages = {L21},
author = {Alexander, K. D. and Berger, E. and Fong, W. and Williams, P. K. G. and Guidorzi, C. and Margutti, R. and Metzger, B. D. and Annis, J. and Blanchard, P. K. and Brout, D. and Brown, D. A. and Chen, H.-Y. and Chornock, R. and Cowperthwaite, P. S. and Drout, M. and Eftekhari, T. and Frieman, J. and Holz, D. E. and Nicholl, M. and Rest, A. and Sako, M. and Soares-Santos, M. and Villar, V. A.},
title = {The Electromagnetic Counterpart of the Binary Neutron Star Merger LIGO/Virgo GW170817. VI. Radio Constraints on a Relativistic Jet and Predictions for Late-time Emission from the Kilonova Ejecta},
journal = {The Astrophysical Journal Letters}
}

@ARTICLE{vista,
       author = {{Sutherland}, Will and {Emerson}, Jim and {Dalton}, Gavin and {Atad-Ettedgui}, Eli and {Beard}, Steven and {Bennett}, Richard and {Bezawada}, Naidu and {Born}, Andrew and {Caldwell}, Martin and {Clark}, Paul and {Craig}, Simon and {Henry}, David and {Jeffers}, Paul and {Little}, Bryan and {McPherson}, Alistair and {Murray}, John and {Stewart}, Malcolm and {Stobie}, Brian and {Terrett}, David and {Ward}, Kim and {Whalley}, Martin and {Woodhouse}, Guy},
        title = "{The Visible and Infrared Survey Telescope for Astronomy (VISTA): Design, technical overview, and performance}",
      journal = {\aap},
         year = 2015,
        month = mar,
       volume = {575},
          eid = {A25},
        pages = {A25},
          doi = {10.1051/0004-6361/201424973},
archivePrefix = {arXiv},
       eprint = {1409.4780},
 primaryClass = {astro-ph.IM},
       adsurl = {https://ui.adsabs.harvard.edu/abs/2015A&A...575A..25S}
}

@ARTICLE{tak14,
       author = {{Takami}, Hajime and {Nozawa}, Takaya and {Ioka}, Kunihito},
        title = "{Dust Formation in Macronovae}",
      journal = {\apjl},
         year = 2014,
        month = jul,
       volume = {789},
       number = {1},
          eid = {L6},
        pages = {L6},
          doi = {10.1088/2041-8205/789/1/L6},
archivePrefix = {arXiv},
       eprint = {1403.5872},
 primaryClass = {astro-ph.HE},
       adsurl = {https://ui.adsabs.harvard.edu/abs/2014ApJ...789L...6T}
}

@ARTICLE{gall17,
       author = {{Gall}, Christa and {Hjorth}, Jens and {Rosswog}, Stephan and {Tanvir}, Nial R. and {Levan}, Andrew J.},
        title = "{Lanthanides or Dust in Kilonovae: Lessons Learned from GW170817}",
      journal = {\apjl},
         year = 2017,
        month = nov,
       volume = {849},
       number = {2},
          eid = {L19},
        pages = {L19},
          doi = {10.3847/2041-8213/aa93f9},
archivePrefix = {arXiv},
       eprint = {1710.05863},
 primaryClass = {astro-ph.HE},
       adsurl = {https://ui.adsabs.harvard.edu/abs/2017ApJ...849L..19G}
}

@ARTICLE{kob21,
  author={Kobyzev, Ivan and Prince, Simon J.D. and Brubaker, Marcus A.},
  journal={IEEE Transactions on Pattern Analysis and Machine Intelligence}, 
  title={Normalizing Flows: An Introduction and Review of Current Methods}, 
  year={2021},
  volume={43},
  number={11},
  pages={3964-3979},
  doi={10.1109/TPAMI.2020.2992934}}

@inproceedings{din17,
title	= {Density estimation using Real NVP},
author	= {Laurent Dinh and Jascha Sohl-Dickstein and Samy Bengio},
booktitle = {International Conference on Learning Representations},
year	= {2017},
URL	= {https://arxiv.org/abs/1605.08803}}

@article{pap19,
  author  = {George Papamakarios and Eric Nalisnick and Danilo Jimenez Rezende and Shakir Mohamed and Balaji Lakshminarayanan},
  title   = {Normalizing Flows for Probabilistic Modeling and Inference},
  journal = {Journal of Machine Learning Research},
  year    = {2021},
  volume  = {22},
  number  = {57},
  pages   = {1--64},
  url     = {http://jmlr.org/papers/v22/19-1028.html}
}

@article{planck2018,
   title={Planck 2018 results: VI. Cosmological parameters},
   volume={641},
   ISSN={1432-0746},
   url={http://dx.doi.org/10.1051/0004-6361/201833910},
   DOI={10.1051/0004-6361/201833910},
   journal={Astronomy and Astrophysics},
   publisher={EDP Sciences},
   author={{Aghanim, N.} and others.},
   year={2020},
   month={Sept}, pages={A6} 
   }

@ARTICLE{thi11_rprocess,
       author = {{Thielemann}, F.-K. and {Arcones}, A. and {K{\"a}ppeli}, R. and {Liebend{\"o}rfer}, M. and {Rauscher}, T. and {Winteler}, C. and {Fr{\"o}hlich}, C. and {Dillmann}, I. and {Fischer}, T. and {Martinez-Pinedo}, G. and {Langanke}, K. and {Farouqi}, K. and {Kratz}, K.-L. and {Panov}, I. and {Korneev}, I.~K.},
        title = "{What are the astrophysical sites for the r-process and the production of heavy elements?}",
      journal = {Progress in Particle and Nuclear Physics},
         year = 2011,
        month = apr,
       volume = {66},
       number = {2},
        pages = {346-353},
          doi = {10.1016/j.ppnp.2011.01.032},
       adsurl = {https://ui.adsabs.harvard.edu/abs/2011PrPNP..66..346T}
}

@article{Dax21realtime,
    author = {Dax, Maximilian and Green, Stephen R. and Gair, Jonathan and Macke, Jakob H. and Buonanno, Alessandra and Sch{\"o}lkopf, Bernhard},
    title = "{Real-Time Gravitational Wave Science with Neural Posterior Estimation}",
    eprint = "2106.12594",
    archivePrefix = "arXiv",
    primaryClass = "gr-qc",
    reportNumber = "LIGO-P2100223",
    doi = "10.1103/PhysRevLett.127.241103",
    journal = "Phys. Rev. Lett.",
    volume = "127",
    number = "24",
    pages = "241103",
    year = "2021"
}

@ARTICLE{Dax25realtime,
       author = {{Dax}, Maximilian and {Green}, Stephen R. and {Gair}, Jonathan and {Gupte}, Nihar and {P{\"u}rrer}, Michael and {Raymond}, Vivien and {Wildberger}, Jonas and {Macke}, Jakob H. and {Buonanno}, Alessandra and {Sch{\"o}lkopf}, Bernhard},
        title = "{Real-time inference for binary neutron star mergers using machine learning}",
      journal = {\nat},
         year = 2025,
        month = mar,
       volume = {639},
       number = {8053},
        pages = {49-53},
          doi = {10.1038/s41586-025-08593-z},
archivePrefix = {arXiv},
       eprint = {2407.09602},
 primaryClass = {gr-qc},
       adsurl = {https://ui.adsabs.harvard.edu/abs/2025Natur.639...49D}
}

@article{Mag20ET,
doi = {10.1088/1475-7516/2020/03/050},
url = {https://doi.org/10.1088/1475-7516/2020/03/050},
year = {2020},
month = {mar},
publisher = {},
volume = {2020},
number = {03},
pages = {050},
author = {Maggiore, Michele and Broeck, Chris Van Den and Bartolo, Nicola and Belgacem, Enis and Bertacca, Daniele and Bizouard, Marie Anne and Branchesi, Marica and Clesse, Sebastien and Foffa, Stefano and García-Bellido, Juan and Grimm, Stefan and Harms, Jan and Hinderer, Tanja and Matarrese, Sabino and Palomba, Cristiano and Peloso, Marco and Ricciardone, Angelo and Sakellariadou, Mairi},
title = {Science case for the Einstein telescope},
journal = {Journal of Cosmology and Astroparticle Physics}
}

@ARTICLE{Eva21CE,
       author = {{Evans}, Matthew and {Adhikari}, Rana X and {Afle}, Chaitanya and {Ballmer}, Stefan W. and {Biscoveanu}, Sylvia and {Borhanian}, Ssohrab and {Brown}, Duncan A. and {Chen}, Yanbei and {Eisenstein}, Robert and {Gruson}, Alexandra and {Gupta}, Anuradha and {Hall}, Evan D. and {Huxford}, Rachael and {Kamai}, Brittany and {Kashyap}, Rahul and {Kissel}, Jeff S. and {Kuns}, Kevin and {Landry}, Philippe and {Lenon}, Amber and {Lovelace}, Geoffrey and {McCuller}, Lee and {Ng}, Ken K.~Y. and {Nitz}, Alexander H. and {Read}, Jocelyn and {Sathyaprakash}, B.~S. and {Shoemaker}, David H. and {Slagmolen}, Bram J.~J. and {Smith}, Joshua R. and {Srivastava}, Varun and {Sun}, Ling and {Vitale}, Salvatore and {Weiss}, Rainer},
        title = "{A Horizon Study for Cosmic Explorer: Science, Observatories, and Community}",
      journal = {arXiv e-prints},
         year = 2021,
        month = sep,
          eid = {arXiv:2109.09882},
        pages = {arXiv:2109.09882},
          doi = {10.48550/arXiv.2109.09882},
archivePrefix = {arXiv},
       eprint = {2109.09882},
 primaryClass = {astro-ph.IM},
       adsurl = {https://ui.adsabs.harvard.edu/abs/2021arXiv210909882E}
}

@article{Dro17,
    author = "Drout, M. R. and others",
    title = "{Light Curves of the Neutron Star Merger GW170817/SSS17a: Implications for R-Process Nucleosynthesis}",
    eprint = "1710.05443",
    archivePrefix = "arXiv",
    primaryClass = "astro-ph.HE",
    doi = "10.1126/science.aaq0049",
    journal = "Science",
    volume = "358",
    pages = "1570--1574",
    year = "2017"
}

@ARTICLE{Nicholl17,
       author = {{Nicholl}, M. and {Berger}, E. and {Kasen}, D. and {Metzger}, B.~D. and {Elias}, J. and {Brice{\~n}o}, C. and {Alexander}, K.~D. and {Blanchard}, P.~K. and {Chornock}, R. and {Cowperthwaite}, P.~S. and {Eftekhari}, T. and {Fong}, W. and {Margutti}, R. and {Villar}, V.~A. and {Williams}, P.~K.~G. and {Brown}, W. and {Annis}, J. and {Bahramian}, A. and {Brout}, D. and {Brown}, D.~A. and {Chen}, H.-Y. and {Clemens}, J.~C. and {Dennihy}, E. and {Dunlap}, B. and {Holz}, D.~E. and {Marchesini}, E. and {Massaro}, F. and {Moskowitz}, N. and {Pelisoli}, I. and {Rest}, A. and {Ricci}, F. and {Sako}, M. and {Soares-Santos}, M. and {Strader}, J.},
        title = "{The Electromagnetic Counterpart of the Binary Neutron Star Merger LIGO/Virgo GW170817. III. Optical and UV Spectra of a Blue Kilonova from Fast Polar Ejecta}",
      journal = {\apjl},
         year = 2017,
        month = oct,
       volume = {848},
       number = {2},
          eid = {L18},
        pages = {L18},
          doi = {10.3847/2041-8213/aa9029},
archivePrefix = {arXiv},
       eprint = {1710.05456},
 primaryClass = {astro-ph.HE},
       adsurl = {https://ui.adsabs.harvard.edu/abs/2017ApJ...848L..18N}
}

@article{Cow17,
doi = {10.3847/2041-8213/aa8fc7},
url = {https://doi.org/10.3847/2041-8213/aa8fc7},
year = {2017},
month = {oct},
publisher = {The American Astronomical Society},
volume = {848},
number = {2},
pages = {L17},
author = {Cowperthwaite, P. S. and Berger, E. and Villar, V. A. and Metzger, B. D. and Nicholl, M. and Chornock, R. and Blanchard, P. K. and Fong, W. and Margutti, R. and Soares-Santos, M. and Alexander, K. D. and Allam, S. and Annis, J. and Brout, D. and Brown, D. A. and Butler, R. E. and Chen, H.-Y. and Diehl, H. T. and Doctor, Z. and Drout, M. R. and Eftekhari, T. and Farr, B. and Finley, D. A. and Foley, R. J. and Frieman, J. A. and Fryer, C. L. and García-Bellido, J. and Gill, M. S. S. and Guillochon, J. and Herner, K. and Holz, D. E. and Kasen, D. and Kessler, R. and Marriner, J. and Matheson, T. and Neilsen, E. H. and Quataert, E. and Palmese, A. and Rest, A. and Sako, M. and Scolnic, D. M. and Smith, N. and Tucker, D. L. and Williams, P. K. G. and Balbinot, E. and Carlin, J. L. and Cook, E. R. and Durret, F. and Li, T. S. and Lopes, P. A. A. and Lourenço, A. C. C. and Marshall, J. L. and Medina, G. E. and Muir, J. and Muñoz, R. R. and Sauseda, M. and Schlegel, D. J. and Secco, L. F. and Vivas, A. K. and Wester, W. and Zenteno, A. and Zhang, Y. and Abbott, T. M. C. and Banerji, M. and Bechtol, K. and Benoit-Lévy, A. and Bertin, E. and Buckley-Geer, E. and Burke, D. L. and Capozzi, D. and Carnero Rosell, A. and Carrasco Kind, M. and Castander, F. J. and Crocce, M. and Cunha, C. E. and D’Andrea, C. B. and Costa, L. N. da and Davis, C. and DePoy, D. L. and Desai, S. and Dietrich, J. P. and Drlica-Wagner, A. and Eifler, T. F. and Evrard, A. E. and Fernandez, E. and Flaugher, B. and Fosalba, P. and Gaztanaga, E. and Gerdes, D. W. and Giannantonio, T. and Goldstein, D. A. and Gruen, D. and Gruendl, R. A. and Gutierrez, G. and Honscheid, K. and Jain, B. and James, D. J. and Jeltema, T. and Johnson, M. W. G. and Johnson, M. D. and Kent, S. and Krause, E. and Kron, R. and Kuehn, K. and Nuropatkin, N. and Lahav, O. and Lima, M. and Lin, H. and Maia, M. A. G. and March, M. and Martini, P. and McMahon, R. G. and Menanteau, F. and Miller, C. J. and Miquel, R. and Mohr, J. J. and Neilsen, E. and Nichol, R. C. and Ogando, R. L. C. and Plazas, A. A. and Roe, N. and Romer, A. K. and Roodman, A. and Rykoff, E. S. and Sanchez, E. and Scarpine, V. and Schindler, R. and Schubnell, M. and Sevilla-Noarbe, I. and Smith, M. and Smith, R. C. and Sobreira, F. and Suchyta, E. and Swanson, M. E. C. and Tarle, G. and Thomas, D. and Thomas, R. C. and Troxel, M. A. and Vikram, V. and Walker, A. R. and Wechsler, R. H. and Weller, J. and Yanny, B. and Zuntz, J.},
title = {The Electromagnetic Counterpart of the Binary Neutron Star Merger LIGO/Virgo GW170817. II. UV, Optical, and Near-infrared Light Curves and Comparison to Kilonova Models},
journal = {The Astrophysical Journal Letters}
}

@ARTICLE{Eva17UV170817,
       author = {{Evans}, P.~A. and {Cenko}, S.~B. and {Kennea}, J.~A. and {Emery}, S.~W.~K. and {Kuin}, N.~P.~M. and {Korobkin}, O. and {Wollaeger}, R.~T. and {Fryer}, C.~L. and {Madsen}, K.~K. and {Harrison}, F.~A. and {Xu}, Y. and {Nakar}, E. and {Hotokezaka}, K. and {Lien}, A. and {Campana}, S. and {Oates}, S.~R. and {Troja}, E. and {Breeveld}, A.~A. and {Marshall}, F.~E. and {Barthelmy}, S.~D. and {Beardmore}, A.~P. and {Burrows}, D.~N. and {Cusumano}, G. and {D'A{\`\i}}, A. and {D'Avanzo}, P. and {D'Elia}, V. and {de Pasquale}, M. and {Even}, W.~P. and {Fontes}, C.~J. and {Forster}, K. and {Garcia}, J. and {Giommi}, P. and {Grefenstette}, B. and {Gronwall}, C. and {Hartmann}, D.~H. and {Heida}, M. and {Hungerford}, A.~L. and {Kasliwal}, M.~M. and {Krimm}, H.~A. and {Levan}, A.~J. and {Malesani}, D. and {Melandri}, A. and {Miyasaka}, H. and {Nousek}, J.~A. and {O'Brien}, P.~T. and {Osborne}, J.~P. and {Pagani}, C. and {Page}, K.~L. and {Palmer}, D.~M. and {Perri}, M. and {Pike}, S. and {Racusin}, J.~L. and {Rosswog}, S. and {Siegel}, M.~H. and {Sakamoto}, T. and {Sbarufatti}, B. and {Tagliaferri}, G. and {Tanvir}, N.~R. and {Tohuvavohu}, A.},
        title = "{Swift and NuSTAR observations of GW170817: Detection of a blue kilonova}",
      journal = {Science},
         year = 2017,
        month = dec,
       volume = {358},
       number = {6370},
        pages = {1565-1570},
          doi = {10.1126/science.aap9580},
archivePrefix = {arXiv},
       eprint = {1710.05437},
 primaryClass = {astro-ph.HE},
       adsurl = {https://ui.adsabs.harvard.edu/abs/2017Sci...358.1565E}
}

@article{Per17,
    author = "Perego, Albino and Radice, David and Bernuzzi, Sebastiano",
    title = "{AT 2017gfo: An Anisotropic and Three-component Kilonova Counterpart of GW170817}",
    eprint = "1711.03982",
    archivePrefix = "arXiv",
    primaryClass = "astro-ph.HE",
    doi = "10.3847/2041-8213/aa9ab9",
    journal = "Astrophys. J. Lett.",
    volume = "850",
    number = "2",
    pages = "L37",
    year = "2017"
}

@article{Pir18,
doi = {10.3847/1538-4357/aaaab3},
url = {https://doi.org/10.3847/1538-4357/aaaab3},
year = {2018},
month = {mar},
publisher = {The American Astronomical Society},
volume = {855},
number = {2},
pages = {103},
author = {Piro, Anthony L. and Kollmeier, Juna A.},
title = {Evidence for Cocoon Emission from the Early Light Curve of SSS17a},
journal = {The Astrophysical Journal}
}

@article{Arc18,
doi = {10.3847/2041-8213/aab267},
url = {https://doi.org/10.3847/2041-8213/aab267},
year = {2018},
month = {mar},
publisher = {The American Astronomical Society},
volume = {855},
number = {2},
pages = {L23},
author = {Arcavi, Iair},
title = {The First Hours of the GW170817 Kilonova and the Importance of Early Optical and Ultraviolet Observations for Constraining Emission Models},
journal = {The Astrophysical Journal Letters}
}

@ARTICLE{Met20,
       author = {{Metzger}, Brian D.},
        title = "{Kilonovae}",
      journal = {Living Reviews in Relativity},
         year = 2020,
        month = dec,
       volume = {23},
       number = {1},
          eid = {1},
        pages = {1},
          doi = {10.1007/s41114-019-0024-0},
archivePrefix = {arXiv},
       eprint = {1910.01617},
 primaryClass = {astro-ph.HE},
       adsurl = {https://ui.adsabs.harvard.edu/abs/2020LRR....23....1M}
}

@article{Gol17GRB,
doi = {10.3847/2041-8213/aa8f41},
url = {https://doi.org/10.3847/2041-8213/aa8f41},
year = {2017},
month = {oct},
publisher = {The American Astronomical Society},
volume = {848},
number = {2},
pages = {L14},
author = {Goldstein, A. and Veres, P. and Burns, E. and Briggs, M. S. and Hamburg, R. and Kocevski, D. and Wilson-Hodge, C. A. and Preece, R. D. and Poolakkil, S. and Roberts, O. J. and Hui, C. M. and Connaughton, V. and Racusin, J. and Kienlin, A. von and Canton, T. Dal and Christensen, N. and Littenberg, T. and Siellez, K. and Blackburn, L. and Broida, J. and Bissaldi, E. and Cleveland, W. H. and Gibby, M. H. and Giles, M. M. and Kippen, R. M. and McBreen, S. and McEnery, J. and Meegan, C. A. and Paciesas, W. S. and Stanbro, M.},
title = {An Ordinary Short Gamma-Ray Burst with Extraordinary Implications: Fermi-GBM Detection of GRB 170817A},
journal = {The Astrophysical Journal Letters}
}

@article{Sav17GRB,
doi = {10.3847/2041-8213/aa8f94},
url = {https://doi.org/10.3847/2041-8213/aa8f94},
year = {2017},
month = {oct},
publisher = {The American Astronomical Society},
volume = {848},
number = {2},
pages = {L15},
author = {Savchenko, V. and Ferrigno, C. and Kuulkers, E. and Bazzano, A. and Bozzo, E. and Brandt, S. and Chenevez, J. and Courvoisier, T. J.-L. and Diehl, R. and Domingo, A. and Hanlon, L. and Jourdain, E. and von Kienlin, A. and Laurent, P. and Lebrun, F. and Lutovinov, A. and Martin-Carrillo, A. and Mereghetti, S. and Natalucci, L. and Rodi, J. and Roques, J.-P. and Sunyaev, R. and Ubertini, P.},
title = {INTEGRAL Detection of the First Prompt Gamma-Ray Signal Coincident with the Gravitational-wave Event GW170817},
journal = {The Astrophysical Journal Letters}
}

@article{cou17,
author = {D. A. Coulter  and R. J. Foley  and C. D. Kilpatrick  and M. R. Drout  and A. L. Piro  and B. J. Shappee  and M. R. Siebert  and J. D. Simon  and N. Ulloa  and D. Kasen  and B. F. Madore  and A. Murguia-Berthier  and Y.-C. Pan  and J. X. Prochaska  and E. Ramirez-Ruiz  and A. Rest  and C. Rojas-Bravo },
title = {Swope Supernova Survey 2017a (SSS17a), the optical counterpart to a gravitational wave source},
journal = {Science},
volume = {358},
number = {6370},
pages = {1556-1558},
year = {2017},
doi = {10.1126/science.aap9811},
URL = {https://www.science.org/doi/abs/10.1126/science.aap9811},
eprint = {https://www.science.org/doi/pdf/10.1126/science.aap9811}}

@article{val17,
doi = {10.3847/2041-8213/aa8edf},
url = {https://doi.org/10.3847/2041-8213/aa8edf},
year = {2017},
month = {oct},
publisher = {The American Astronomical Society},
volume = {848},
number = {2},
pages = {L24},
author = {Valenti, Stefano and Sand, David, J. and Yang, Sheng and Cappellaro, Enrico and Tartaglia, Leonardo and Corsi, Alessandra and Jha, Saurabh W. and Reichart, Daniel E. and Haislip, Joshua and Kouprianov, Vladimir},
title = {The Discovery of the Electromagnetic Counterpart of GW170817: Kilonova AT 2017gfo/DLT17ck},
journal = {The Astrophysical Journal Letters}
}

@ARTICLE{arc17,
       author = {{Arcavi}, Iair and {Hosseinzadeh}, Griffin and {Howell}, D. Andrew and {McCully}, Curtis and {Poznanski}, Dovi and {Kasen}, Daniel and {Barnes}, Jennifer and {Zaltzman}, Michael and {Vasylyev}, Sergiy and {Maoz}, Dan and {Valenti}, Stefano},
        title = "{Optical emission from a kilonova following a gravitational-wave-detected neutron-star merger}",
      journal = {\nat},
         year = 2017,
        month = nov,
       volume = {551},
       number = {7678},
        pages = {64-66},
          doi = {10.1038/nature24291},
archivePrefix = {arXiv},
       eprint = {1710.05843},
 primaryClass = {astro-ph.HE},
       adsurl = {https://ui.adsabs.harvard.edu/abs/2017Natur.551...64A}
}

@article{hu17,
title = {Optical observations of LIGO source GW 170817 by the Antarctic Survey Telescopes at Dome A, Antarctica},
journal = {Science Bulletin},
volume = {62},
number = {21},
pages = {1433-1438},
year = {2017},
issn = {2095-9273},
doi = {https://doi.org/10.1016/j.scib.2017.10.006},
url = {https://www.sciencedirect.com/science/article/pii/S2095927317305224},
author = {Lei Hu and Xuefeng Wu and Igor Andreoni and Michael C. {B. Ashley} and Jeff Cooke and Xiangqun Cui and Fujia Du and Zigao Dai and Bozhong Gu and Yi Hu and Haiping Lu and Xiaoyan Li and Zhengyang Li and Ensi Liang and Liangduan Liu and Bin Ma and Zhaohui Shang and Tianrui Sun and N.B. Suntzeff and Charling Tao and Syed A. Uddin and Lifan Wang and Xiaofeng Wang and Haikun Wen and Di Xiao and Jin Xu and Ji Yang and Shihai Yang and Xiangyan Yuan and Hongyan Zhou and Hui Zhang and Jilin Zhou and Zonghong Zhu}
}

@Article{matplotlib,
  Author    = {Hunter, J. D.},
  Title     = {Matplotlib: A 2D graphics environment},
  Journal   = {Computing in Science \& Engineering},
  Volume    = {9},
  Number    = {3},
  Pages     = {90--95},
  publisher = {IEEE COMPUTER SOC},
  doi       = {10.1109/MCSE.2007.55},
  year      = 2007
}
\bibliographystyle{aasjournalv7}

\end{CJK*}
\end{document}